\documentclass[aps, prd,floats, floatfix, twocolumn,superscriptaddress, nofootinbib]{revtex4-1}

\usepackage{graphicx}
\usepackage{hyperref}
\usepackage{amsmath,amssymb}
\usepackage{amsfonts}
\usepackage{xspace} % Sensible space treatment at end of simple macros
\usepackage[usenames]{color}
\usepackage{dcolumn} % Align table columns on decimal point
\usepackage{bm} % bold math
\usepackage{epsfig}

\definecolor {darkgreen}{rgb}{0.2,0.7,0.2}

\newcommand\be{\begin{equation}}
\newcommand\ba{\begin{eqnarray}}
\newcommand\ee{\end{equation}}
\newcommand\ea{\end{eqnarray}}

\newcommand{\GW}{{\mbox{\tiny GW}}}
\newcommand{\Ho}{{\mbox{\tiny Hor}}}

\newcommand{\Newt}{{\mbox{\tiny Newt}}}

\newcommand{\Ta}{{\mbox{\tiny Tay.}}}

\newcommand{\NS}{{\mbox{\tiny NS}}}
\newcommand{\Sp}{{\mbox{\tiny S}}}
\newcommand{\ISCO}{{\mbox{\tiny ISCO}}}

\begin{document}

\title{Extreme Mass-Ratio Inspirals in the Effective-One-Body Approach: 
\\ Quasi-Circular, Equatorial Orbits around a Spinning Black Hole}

\author{Nicol\'as Yunes}
\affiliation{Department of Physics, Princeton University, Princeton, NJ 08544}
\affiliation{Department of Physics and MIT Kavli Institute, 77 Massachusetts Avenue, Cambridge, MA 02139}
\affiliation{Harvard-Smithsonian, Center for Astrophysics, 60 Garden St., Cambridge, MA 02138, USA.}

\author{Alessandra~Buonanno}
\affiliation{Maryland Center for Fundamental Physics \& Joint Space-Science Institute,  
Department of Physics, University of Maryland, College Park, MD 20742}

\author{Scott A.~Hughes}
\affiliation{Department of Physics and MIT Kavli Institute, 77 Massachusetts Avenue, Cambridge, MA 02139}

\author{Yi~Pan}
\affiliation{Maryland Center for Fundamental Physics \& Joint Space-Science Institute,  
Department of Physics, University of Maryland, College Park, MD 20742}
%\affiliation{Maryland Center for Fundamental Physics, Department of Physics, University of Maryland, College Park, MD 20742}

\author{Enrico~Barausse}
\affiliation{Maryland Center for Fundamental Physics \& Joint Space-Science Institute,  
Department of Physics, University of Maryland, College Park, MD 20742}
%\affiliation{Maryland Center for Fundamental Physics, Department of Physics, University of Maryland, College Park, MD 20742}

\author{M.~Coleman~Miller}
\affiliation{Maryland Astronomy Center for Theory and Computation \& Joint Space-Science Institute, 
 Department of Astronomy, University of Maryland, College Park, MD 20742}

\author{William~Throwe}
\affiliation{Department of Physics and MIT Kavli Institute, 77 Massachusetts Avenue, Cambridge, MA 02139}

%-----------------------------------------------------------------------------
\begin{abstract}

We construct effective-one-body waveform models suitable for data analysis with LISA 
for extreme-mass ratio inspirals in quasi-circular, equatorial orbits about a 
spinning supermassive black hole. 
The accuracy of our model is established through comparisons against
frequency-domain, Teukolsky-based waveforms in the radiative
approximation.
The calibration of eight high-order post-Newtonian parameters in the 
energy flux suffices to obtain a phase and fractional amplitude agreement of better 
than $1$ radian and $1 \%$ respectively over a period between $2$ and $6$ months
depending on the system considered. This agreement translates into matches higher 
than $97\,\%$ over a period between $4$ and $9$ months, depending on the system. 
Better agreements can be obtained if a larger number of calibration parameters 
are included.
Higher-order mass ratio terms in the effective-one-body
Hamiltonian and radiation-reaction introduce phase corrections
of at most $30$ radians in a one year evolution. 
These corrections are usually  
one order of magnitude larger than those introduced by the spin of 
the small object in a one year evolution. 
These results suggest that the effective-one-body approach for extreme mass ratio 
inspirals is a good compromise between accuracy and computational price for
LISA data analysis purposes. 

\end{abstract}

\date{\today \hspace{0.2truecm}}

\maketitle

%-----------------------------------------------------------------------------
\section{Introduction}

Extreme mass-ratio inspirals (EMRIs) are one of the most promising
sources of gravitational waves (GWs) expected to be detected with
the proposed Laser Interferometer Space Antenna (LISA)~
\cite{Bender:1998,Danzmann:2003tv,Prince:2003aa,2004CQGra..21S1595G}.
These sources consist of a small compact object, such as a
neutron star or stellar-mass black hole (BH), in a close orbit
around a spinning, supermassive BH~\cite{2007CQGra..24..113A}.
Gravitational radiation losses cause the small object to spiral
closer to the supermassive BH and eventually merge with it.
Hence, the GW signal from such events encodes information about strong
gravity, allowing tests of general relativity~\cite{Sopuerta:2009iy} 
and of the Kerr metric~\cite{Collins:2004ex,Vigeland:2009pr,Hughes:2006pm,Apostolatos:2009vu,LukesGerakopoulos:2010rc,Glampedakis:2005cf,2008PhRvD..77b4035G,Barausse:2007dy,Barausse:2006vt,Kesden:2004qx,BC}, as well as measurements
of the spins and masses of massive BHs~\cite{Barack:2003fp}.

Unfortunately, EMRIs are very weak sources of GWs at their expected distances from us,
and thus, they must be observed over many cycles to be detectable~\cite{2007CQGra..24..113A}.
For example, a typical
EMRI at a distance of $3 \, \rm{Gpc}$ would produce GWs with signal-to-noise ratios (SNRs) 
on the order of $10$-$200$ depending on the observation time. Therefore, matched
filtering is essential to extract EMRIs
from LISA noise and the foreground of unresolved GWs 
from white dwarf binaries in our galaxy.  

Matched filtering consists of cross-correlating the data stream with
a certain noise-weighted waveform template~\cite{lrr-2005-3}. If the
latter is similar to a GW event hidden in the data, then this
cross-correlation filters it out of the noise. Of course, for
matched filtering to be effective, one must construct accurate
template filters. Otherwise, real events can be missed, or if an
event is detected, parameter estimation can be strongly
biased~\cite{Cutler:2007mi}. The construction of accurate EMRI
waveforms is extremely difficult due to the long duration of the
signal and the strong-field nature of the orbits. A one-year EMRI
signal contains millions of radians in phase information. 
To avoid significant dephasing, its waveform modeling must be 
accurate to at least one part in
$10^{5}$--$10^{6}$~\cite{Lindblom:2008cm}.

% Describe Complications and Simplifications that apply to EMRIs. 
% include the following: 
%	- inspiral can reach velocities of 0.7 the speed of light -> PN not very good here unless resummed/fitted.
%	- BH perturbation theory is ideal, but computational cost too high + need for second order/post-adiabatic?
Such an exquisite accuracy requirement is complicated further by the
strong-field nature of the orbit. An EMRI can reach orbital
velocities of two-thirds the speed of light and orbital separations
as small as a few times the mass of the supermassive companion. This
automatically implies that standard, post-Newtonian (PN)
Taylor-expanded waveforms fail to model such EMRI
orbits~\cite{Mandel:2008bc}. PN theory relies on the assumptions
that all orbital velocities are much smaller than the speed of light
and that all objects are at separations much larger than the total
mass of the system~\cite{Blanchet:2002av}. A better approximation
scheme to model EMRIs is BH perturbation theory, where one only
assumes that the mass ratio of the system is much less than
unity~\cite{Mino:1997bx}. This is clearly the case for EMRIs, as the
mass ratio is in the range $10^{-4}\mbox{--}10^{-6}$. Perturbation theory,
however, is computationally and analytically expensive. Only
recently have generic orbits been computed around a non-spinning BH
to linear order in the mass ratio~\cite{Barack:2009ey,Sago:2009zz}, and it is unlikely that these
will be directly used for EMRI data analysis~\cite{2004CQGra..21S1595G}.

%	- merger and ringdown can be neglected. 
EMRIs involve complicated inspiral analysis, but unlike 
comparable-mass coalescences, the merger and ringdown phase can
be completely neglected. To see this, note that the instantaneous amplitude
of the waves from a binary scales as $\mu$, where $\mu =m_{1}\, m_{2}/M$ is
the reduced mass, $M$ is the total mass and $m_{1,2}$ are the component masses.
The inspiral lasts for a time $\sim 1/\mu$ and releases an energy flux $\sim \mu^2/\mu\sim \mu$.
In contrast, the merger and ringdown last for a time $\sim M$
(independent of $\mu$), and thus, release an energy flux $\sim \mu^2$.  For
an EMRI, $\mu\ll M$ and the inspiral clearly dominates the signal.
Based on this argument, we neglect the merger and ringdown, focusing on the inspiral for our analysis.

\subsection{Summary of Previous Work} 
% Describe what we intend to do here: Combine EOB with BH Perturbation theory. 
% Describe Teuk Modeling and EOB Modeling
The modeling of EMRIs has been attempted in the past with various 
degrees of success. One approach is to compute the self-field of the 
test particle to understand how it modifies the orbital 
trajectory. This task, however, is quite involved, both theoretically 
and computationally, as the self-field contains a divergent piece that 
is difficult to regularize (see, e.g. Ref.~\cite{Barack:2009ux} for a recent 
review). Recently, a breakthrough was achieved, with the full 
calculation of the self-force for generic EMRIs about non-spinning 
supermassive BHs~\cite{Barack:2009ey,Sago:2009zz}. Such calculations, 
however, are computationally prohibitive if the goal is to populate a 
waveform template space.

Another approach is to use more approximate methods to model the EMRI 
trajectories. One such approach was developed by 
Hughes~\cite{Hughes:1999bq,Hughes:2001jr}, following the pioneering work 
of Poisson~\cite{Poisson:1993vp}. In this {\emph{radiative-adiabatic}} 
scheme, the inspiral is treated as a sequence of adiabatically shrinking 
geodesics. The degree of shrinkage is determined by solving the 
Teukolsky equation on each individual geodesic. Its solution encodes
how the {\it constants} of the motion (the energy, angular momentum and Carter 
constant) change due to GW emission. By interpolating across such 
sequence of geodesics, one then obtains a continuous inspiral and 
waveform. The calculation of a single waveform, however, is rather 
computationally expensive, as it requires the mapping of the entire 
orbital phase space, which for generic orbits is likely to be prohibitive.
It is also worth noting that the radiative approximation neglects the impact 
of conservative effects which, especially for eccentric orbits, are likely to be 
important~\cite{2005PhRvD..72l4001P}.

Other, perhaps more rough approximations can also be used to model 
EMRIs. The templates obtained through these methods are sometimes called 
{\it kludge} waveforms to emphasize their approximate nature. The goal of 
their construction was never to provide sufficiently accurate templates 
for real data analysis. Instead, kludge waveforms were built to carry 
out {\emph{descoping}} or parameter estimation studies to determine 
roughly the accuracy to which parameters could be extracted, given an 
EMRI detection with LISA.

The first kludge waveforms were constructed by Barack and 
Cutler~\cite{Barack:2003fp}. These waveforms employ the quadrupole 
formula to build templates as a function of the orbital trajectories. 
The latter are simply Keplerian ellipses with varying orbital elements. 
The variation of these is determined by low-order PN expressions, 
constructed from the GW energy and angular momentum fluxes. An 
improvement of these fluxes was developed by Gair and 
Glampedakis~\cite{Gair:2005ih}, who fitted these low-order PN expression 
to more accurate fluxes constructed from solutions to the Teukolsky 
equation. A further improvement was developed by Babak 
et~al.~\cite{Babak:2006uv}, who modeled the waveforms via a 
quadrupole-octopole formula and the orbital trajectories via solutions 
to the geodesic equations, augmented with PN--orbit-averaged evolution 
equations for the orbital elements.

All of these improvements, however, do not mean that kludge waveforms 
would be effectual or faithful for realistic data analysis with LISA. 
One cannot exactly quantify this statement because exact EMRI waveforms 
are not available and will not be in the near future. One can 
nonetheless predict that these approaches will be insufficient because 
critical components of the fluxes are not being taken into account. For 
example, GWs do not only escape to infinity, but they are also absorbed 
by the supermassive BH, contributing to the overall fluxes of energy and 
angular momentum. This contribution is non-negligible if one considers 
sufficiently long waveforms (longer than a few weeks). In fact, as we 
shall show in this paper, even the inclusion of such terms and very high 
order PN expressions in the fluxes is still insufficient for accurate 
waveform models that last more than a couple of months.

\subsection{The Effective-One-Body Approach} 

% Previous EOB work
The effective-one-body (EOB) formalism was introduced in 
Refs.~\cite{Buonanno99,Buonanno00} to model the inspiral, merger, and
ringdown of comparable-mass BH binaries. This scheme was then
extended to higher PN orders~\cite{Damour00}, spinning
BHs~\cite{Damour01,Buonanno06,Damour:2008qf,Barausse:2009xi}, small 
mass-ratio mergers~\cite{Nagar:2006xv,Damour2007,Bernuzzi:2010ty}, and 
improved by resumming the radiation reaction-force and waveforms
~\cite{Damour:1997ub,Damour2007,Damour:2008gu,Pan:2010hz,Fujita:2010xj}. In the comparable
mass case, phase and amplitude agreement was achieved between EOB
and numerical-relativity waveforms, after calibrating a few
parameters~\cite{Damour:2009kr,Buonanno:2009qa,Pan:2009wj}. 
By calibrating the EOB model to the comparable mass case, one 
can also improve the agreement of the model with the self-force predictions~
\cite{Barack:2009ey,Damour:2009sm}. The combination of EOB and BH perturbation 
theory tools for LISA data-analysis purposes 
was first carried out in Refs.~\cite{Yunes:2009ef,AmaroSeoane:2010ub}. In these papers, 
the EOB scheme was found successful for the coherent modeling of EMRIs
about a non-spinning background for a $2$ year period. Here we
extend these results to non-precessing EMRIs about a spinning
background.

% Why do we pick this system
% Astrophysical Motivation
As a first step toward the construction of accurate EMRI waveforms,
we concentrate on quasi-circular, equatorial EMRIs about a
spinning, supermassive Kerr BHs. The modeling of such EMRIs is
simpler than that of inclined and eccentric ones, as only a single
component of the radiation-reaction force is non-vanishing and
entirely controlled by the GW energy flux (the Carter constant
vanishes by symmetry). Moreover, such EMRIs are expected in at
least one astrophysical scenario~\cite{Levin:2006uc}. In this setup, 
stellar-mass compact objects are either created in the accretion
disk surrounding the supermassive BH or are captured by the
disk, and hence move with the disk. The accretion disk is expected 
to be in the spin equatorial plane within a few hundred gravitational radii
of the supermassive BH~\cite{1975ApJ...195L..65B}.

% Present results 1: Fitting of EOB to Teuk highlights
We first calibrate the EOB energy flux to the leading-order energy
flux computed in BH perturbation theory through the solution to
the Teukolsky equation. This calibration is more complicated than
for non-spinning systems because it must now be performed
globally, i.e.,~as a function of both spin and velocity. This
increases the computational cost of the calibration and the number
of calibration parameters, as a bivariate series generically
contains more terms than a monovariate one. After calibrating $8$
parameters, we find that the fluxes agree to within one part in
$10^{3}$ for all spins [$a/M=(-0.99,0.99)$] and velocities [$v=(0.01\,c,v_{\ISCO})$] 
considered, where $v_{\ISCO}$ is the velocity at the innermost circular orbit (ISCO). 

% Present results 2: Comparison of EOB and Teuk phases and amplitudes
Once the energy flux has been calibrated, we evolve the Hamilton equations
in the adiabatic approximation and compare the amplitude and phase
evolution to that obtained with an approximate BH perturbation
theory, numerical result. For the latter, we employ the so-called
radiative approximation~\cite{Hughes:1999bq,Hughes:2001jr}, where
one models the EMRI as an adiabatic sequence of 
geodesics with varying orbital elements, as prescribed by the
solution to the Teukolsky equation. We find that the EOB and
Teukolsky-based waveforms agree in phase and relative amplitude to
better than $1$ radian and $1 \%$ respectively after $2$ or $6$ months of
evolution, depending on the system considered. Better agreements can be 
obtained if a larger number of calibration parameters were included.

%How our work differs from previous BHPT one. 
Our EOB waveforms differ from previous kludge models on several fronts.
First, the radiation-reaction force is here computed differently than in the kludge approach. 
In the latter this force is calculated from PN, Taylor-expanded fluxes that encode 
the GW that escape to infinity only. These fluxes were then improved by fitting a very 
large number of parameters to more accurate Teukolsky-fluxes with a log-independent, power-series expansion 
for the fitting functions~\cite{Gair:2005ih}. In the EOB approach, the radiation-reaction force is 
computed directly from the factorized resummed waveforms~\cite{Damour:2008gu,Pan:2010hz}. 
These are enhanced through the addition of BH absorption terms and then the calibration of eight high PN-order parameters to Teukolsky fluxes with a log-dependent, power-series expansion for the fitting functions. 
Second, the conservative dynamics are also treated here differently than in the kludge approach. 
In the latter, the Hamiltonian is either a two-body, Newtonian one~\cite{Barack:2003fp} or the full 
test-particle limit one, i.e.,~Schwarzschild or Kerr~\cite{Babak:2006uv}. In the EOB approach, 
the conservative dynamics not only encodes the exact test-particle limit Hamiltonian, but they also
allow for the inclusion of finite mass-ratio terms and of the spin of the small body. 

\subsection{Data Analysis Implications} 

The waveforms computed here are thus suitable for coherent data analysis over periods of several months.
This can be established by computing the overlap between the EOB and Teukolsky-based
waveforms, after maximizing over extrinsic parameters (an overall phase and time shift).
We find that, when eight calibration parameters are used, the overlap remains higher than $97 \, \%$ 
over $4$ to $9$ months of evolution, depending on the system considered. This is to be compared with numerical kludge 
waveforms~\cite{Babak:2006uv} whose overlap drops to $56  \, \%$ 
and $74 \, \%$ after $4$ and $9$ months
respectively, even when forty-five calibration parameters are used to fit the flux~\cite{Gair:2005ih} . 
Of course, one could obtain higher overlaps by maximizing over intrinsic parameters, such as the chirp mass 
or the spin of the background, but this would naturally bias parameter estimation. Also, when integrating over only two weeks, the overlap increases, remaining higher than $0.99999$ at $1$ Gpc regardless of the model used. 

The benefit of coherently integrating over longer periods of time is that the recovered SNR 
naturally increases, thus allowing us to detect signals farther out and improving parameter estimation. 
One can see this by simply noting that the SNR scales with the square root of the
time of observation. For example, coherent integration over $4$ or $9$ months instead of two weeks 
increases the SNR at $1$ Gpc from $13$ to $64$ and from $27$ to $85$ for two prototypical EMRIs. 
Such a large increase in SNR by coherently integrating over long observation times brings 
EMRIs not only to a confidently detectable range, but would also allow interesting tests of GR. 

We conclude the paper by studying the error introduced in these waveforms due
to neglecting second-order mass-ratio terms in the radiation-reaction force
(dissipative PN self-force) and first-order in the mass-ratio terms in the Hamiltonian 
(conservative PN self-force). Such mass-ratio dependent effects can easily be included in the EOB
prescription, as they are known in the PN/EOB framework. Of course,
since these are known to finite PN order, we cannot include full
second-order effects. These effects should be considered estimates, 
since the complete result may differ from the PN prediction. 
We find that such PN radiation-reaction
effects modify the phase of the waveform by ${\cal{O}}(10)$ radians in a one
year evolution, provided the EMRI samples the strong-gravity regime close to the ISCO. 
In a two-month period, however, the inclusion of finite mass-ratio effects increases the mismatch
from $2.9 \times 10^{-5}$ to $3.6 \times 10^{-5}$ at 1 Gpc. This implies that such effects
will only be seen if one coherently integrates over a sufficiently long time of observation.
We find a somewhat smaller final dephasing when we allow the
second body to be spinning and neglect any self-force corrections. The relative importance 
of the conservative or dissipative PN self-force terms and that of the spin of the second body 
depends on somewhat on the EMRI considered. Generically, we find all such corrections
to be larger than one radian after a full year of integration, while they are negligible over a two month period. 

% Paper Organization and conventions used.
This paper is organized as follows. 
Section~\ref{sec:modeling} describes how we model EMRIs analytically and numerically. 
Section~\ref{sec:calibration} discusses how the analytical EOB model is calibrated to the Teukolsky energy flux. 
Section~\ref{sec:results} compares EOB evolutions to Teukolsky ones, while 
Sec.~\ref{sec:DA} discusses the data analysis implications of such a comparison. 
Section~\ref{sec:hi-order} estimates the effect of mass-ratio dependent effects and
Sec.~\ref{sec:conclusions} concludes and points to future research. 
Appendix~\ref{app:sph} presents details on the transformation between spheroidal
and spherical tensor harmonics. Appendix~\ref{app:flux-abs} contains expressions for the GW energy flux absorbed by BHs. Finally, in Appendix~\ref{app:seob} we write the EOB Hamiltonian derived in Ref.~\cite{Barausse:2009xi} 
when BHs carry spins aligned or anti-aligned with the orbital angular momentum. 
We use geometric units, with $G = c = 1$, unless otherwise noted. 

\section{EMRI modeling}
\label{sec:modeling}

\subsection{Analytical modeling: EOB-based waveforms}
\label{sec:Amodeling}

Consider a BH binary system with masses $m_{1}$ and $m_{2}$, total mass $M = m_{1} + m_{2}$, 
reduced mass $\mu = m_{1} m_{2}/M$ and symmetric mass ratio $\nu = \mu/M$. 
We assume that the orbital angular momentum is co-aligned or counter-aligned with the individual BH spins
$S_{A} = a_{A} m_{A}=q_A m_A^2$, where $a_{A}=S_{A}/m_A$ denotes the $A$th BH's spin parameter
and $q_{A} = a_{A}/m_{A}$ denotes the dimensionless spin parameter.

We first discuss the case of a non-spinning BH ($q_{2} = 0$) with mass $m_2$ 
orbiting a spinning BH with spin parameter $q_{1}$ and mass $m_1\gg m_2$, to leading
order in the mass ratio $m_{2}/m_{1}$. Subleading terms in the mass ratio and terms 
proportional to $q_{2}$ introduce conservative corrections that are not included in the Teukolsky waveforms, 
which we shall use to calibrate our model, and we therefore neglect them during the calibration. Eventually, 
however, we shall turn these conservative terms on and estimate their effect using the spin EOB 
Hamiltonian of Ref.~\cite{Barausse:2009xi} (see Appendix~\ref{app:seob}). 

In the EOB framework, the orbital trajectories are obtained by solving Hamilton's equations, supplemented by 
a radiation-reaction force describing the backreaction of GW emission on the orbital dynamics. 
Neglecting conservative corrections of order ${\cal O}(m_2/m_1)$ and ${\cal O}(q_2)$, the spin 
EOB Hamiltonian reduces to the Hamiltonian of a non-spinning test-particle 
in Kerr, $H_{\rm NS}$:
\begin{gather}
{H}_{\rm EOB} = {H}_{\rm NS}\, [1+{\cal O}(m_2/m_1)+{\cal O}(q_2)]\,,\\ 
{H}_{\rm NS} = \beta^i \, p_i + \alpha \sqrt{m_2^2 + \gamma^{ij}\,p_i\,p_j}\,,
\label{eq:Hnsdef}
\end{gather} 
where 
\begin{eqnarray}
\label{alpha}
\alpha &=& \frac{1}{\sqrt{-g^{tt}}}\,,\\
\beta^i &=& \frac{g^{ti}}{g^{tt}}\,,\\
\gamma^{ij} &=& g^{ij}-\frac{g^{ti}\,g^{tj}}{g^{tt}}\,,
\label{gamma}
\end{eqnarray}
$g_{\mu\nu}$ being the Kerr metric. In Boyer-Lindquist coordinates $(t,r,\theta,\phi)$ and restricting 
ourselves to the equatorial plane $\theta = 0$, the relevant metric components read
\begin{subequations}
\begin{eqnarray}
\label{def_metric_in}
g^{tt} &=& -\frac{\Lambda}{r^2\,\Delta}\,,\\
g^{rr} &=& \frac{\Delta}{r^2}\,,\\
g^{\phi\phi} &=& \frac{1}{\Lambda}
\left(-\frac{{\omega}_{\rm fd}^2}{r^2\,\Delta} + r^2\right)\,,\label{eq:gff}\\
g^{t\phi}&=&-\frac{{\omega}_{\rm fd}}{r^2\,\Delta}\,,\label{def_metric_fin}
\end{eqnarray}
\end{subequations}
where ${\omega}_{\rm fd}= 2 q_{1}\, m_1^2\, r$, and the metric potentials are
\begin{eqnarray}
\label{Delta}
\Delta &=& r^2 + q_{1}^2\,m_1^2 - 2m_1\,r \,, \\
\Lambda &=& (r^2 + q_{1}^2\,m_1^2)^2 - q_{1}^2\,m_1^2\,\Delta\,.
\label{Lambda}
\end{eqnarray}
Although the EOB formalism includes possible 
non-adiabaticities in the last stages of the inspiral and plunge, it is not necessary 
to include non-adiabatic effects here. Generically, for the systems that we consider, 
we find that the inclusion of non-adiabatic corrections leads to small phase corrections 
(of ${\cal{O}}(1 \; {\rm{rad}})$ after one year of evolution)~\cite{Yunes:2009ef,AmaroSeoane:2010ub}. 
The assumption of adiabaticity allows us to simplify the evolution equations that are solved 
numerically. This in turn reduces the computational cost of producing EOB waveforms: an adiabatic EOB evolution requires a few CPU seconds, while a non-adiabatic one would require CPU days or weeks.
The non-adiabatic model is computationally more expensive because one needs to solve all of Hamilton's equations, with radiation reaction source terms that are expensive to evaluate.  

The Hamiltonian of Eq.~(\ref{eq:Hnsdef}) simplifies drastically when we consider circular, equatorial orbits ($\theta = \pi/2$)
with $\mathbf{S}_{1}$ co-aligned or counter-aligned with the orbital angular momentum (see eg.~\cite{Bardeen:1972fi}).
Imposing $p_r = 0$, which is a necessary condition for circular orbits, and inserting Eqs.~(\ref{def_metric_in})--(\ref{def_metric_fin}) 
in Eq.~(\ref{eq:Hnsdef}), a straightforward calculation returns
\begin{equation}
\label{HKerr}
H_{\rm NS}=p_{\phi}\,\frac{{\omega}_{\rm fd}}{\Lambda}+ \frac{m_2\,r\,\sqrt{\Delta}\,\sqrt{Q}}{\sqrt{\Lambda}}\,,
\end{equation}
where
\begin{equation}
Q=1+\frac{p_\phi^2\, r^2}{m_2^2\, \Lambda}\,,
\end{equation}
and $p_{\phi} \equiv L$ is the conjugate momentum to the $\phi$ Boyer-Lindquist coordinate or simply the orbital
angular momentum. Imposing the condition $\dot{p}_r = (\partial H_{\rm NS}/\partial r)_{p_r=0} = 0$, which is also satisfied by circular orbits, 
we can solve for $L$ as a function of $r$ and $q_{1}$~\cite{Bardeen:1972fi} 
\begin{equation}
L= \pm m_2\,m_1^{1/2}\,\frac{r^2 \mp 2 q_{1}\,m_1^{3/2}\,r^{1/2} + q_{1}^2\,m_1^2}{r^{3/4} \left(r^{3/2} - 3 m_1 r^{1/2} \pm 2 q_{1}\,m_1^{3/2}\right)^{1/2}}\,,
\label{ang-mom}
\end{equation}
where $\pm$ corresponds to prograde or retrograde orbits, respectively. 
Inserting the above equation in Eq.~(\ref{HKerr}) yields an expression for the energy $E \equiv 
H^{\rm circ}_{\rm NS}(r)$ of circular orbits in Kerr~\cite{Bardeen:1972fi} 
\begin{equation}
E=m_{1} + m_2\,\frac{1-2m_1/r \pm q_{1}\,m_1^{3/2}/r^{3/2}}{\sqrt{1-3m_1/r\pm 2q_{1}\,m_1^{3/2}/r^{3/2}}}\,.
\end{equation}
The above quantities can also be expressed in terms of the orbital velocity $\omega$, once $r(\omega)$ 
is derived for circular orbits. Computing $\omega = (\partial H_{\rm NS}/\partial p_\phi)_{p_r=0}$, 
using Eq.~(\ref{ang-mom}), we obtain
\begin{equation}
r = \frac{\left[1 - q_{1}\, (m_1 \omega) \right]^{2/3}}{(m_1 \omega)^{2/3}}\,.
\label{r-of-w}
\end{equation}
We also define the parameter $v \equiv (m_{1} \omega)^{1/3}$.

In the adiabatic approximation, the orbital evolution is fully determined by the frequency evolution through Eq.~\eqref{r-of-w}.  
Assuming the motion follows an adiabatic sequence of quasi-circular orbits, we can use 
the balance equation $\dot{L} = \dot{E}/\omega = - {\cal F}/\omega$ to derive
\begin{equation}
\label{omegaadiab}
\dot{\omega} = - \frac{1}{\omega} \left(\frac{d L}{d \omega}\right)^{-1} \, {\cal F}(\omega)\,,
\end{equation}
where ${\cal F}$ is the GW energy flux (see e.g.~Ref.~\cite{Buonanno00}). The multipolar 
factorized form of this flux, proposed in the non-spinning case in Refs.~\cite{Damour2007,Damour:2008gu} and extended 
to the spin case in Ref.~\cite{Pan:2010hz}, is given by
\begin{eqnarray}
{\cal F}(\omega) \equiv \frac{1}{8 \pi}\,\sum_{\ell=2}^{8}\sum_{m=0}^{\ell} \left|\dot{h}_{\ell m} \right|^{2}\,,
\label{pd-flux}
\end{eqnarray}
which under the assumption of adiabaticity reduces to
\begin{eqnarray}
{\cal F}(\omega)= \frac{1}{8 \pi}\,\sum_{\ell=2}^{8}\sum_{m=0}^{\ell} (m\,\omega)^2\,\left|h_{\ell m} \right|^{2}\,,
\label{pd-flux-adiab}
\end{eqnarray}
with 
\begin{equation}
h_{\ell m}(v) = h_{\ell m}^{\Newt,\epsilon_p}\;S^{\epsilon_p}_{\ell m} \; T_{\ell m}\; e^{i \delta_{\ell m}}\; (\rho_{\ell m})^\ell\,,
\label{full-h}
\end{equation}
where $\epsilon_p$ denotes the parity of the multipolar waveform (\textit{i.e.,} $\epsilon_p=0$ if $\ell+ m$ is even, $\epsilon_p=1$ if $\ell+ m$ is odd), and 
\begin{equation}
h_{\ell m}^{\Newt,\epsilon_p} \equiv \frac{m_1}{R}\, n^{(\epsilon_p)}_{\ell m}\, c_{\ell +\epsilon_p}\, v^{\ell +\epsilon_p}\, 
Y_{\ell - \epsilon_p,-m}(\pi/2,\phi).
\end{equation}
When spin effects are present, the expressions for all the terms in
Eq.~\eqref{full-h}, namely $S^{\epsilon_p}_{\ell m}(v)$, $T_{\ell m}(v)$,
$\delta_{\ell m}(v)$ and $\rho_{\ell m}(v)$ can be read in 
Ref.~\cite{Pan:2010hz} [see Eqs.~(24), (25), (26) and (29)
therein]. The functions $Y_{\ell,m}(\theta,\phi)$ are the standard
spherical harmonics, while $n_{\ell m}^{(\epsilon_p)}$ and $c_{\ell
  +\epsilon_p}$ are numerical coefficients that depend on the mass
ratio (see Eqs.~$(5)$-$(7)$ in Ref.~\cite{Damour:2008gu}). As before,
we work to leading order in $\nu$ initially, and later study how the
terms of higher-order in $\nu$ affect the GW phase evolution.

The solution to Eq.~(\ref{omegaadiab}) requires that we prescribe
initial data. We here choose post-circular initial conditions, as
described in Ref.~\cite{Buonanno00}, to set-up a {\it mock} evolution
that starts at a separation of $100 m_1$ and ends at either the ISCO
or whenever the GW frequency reaches $0.01 \, {\rm{Hz}}$. This mock
evolution is then used to read initial data one-year before the end of
the mock evolution. This approach leads to an accurate initial data
prescription, without any eccentricity contamination. For example, the
error in the initial frequency induced by starting the mock evolution
at $100 \, m_1$, instead of $200 \, m_1$, is on the order of $10^{-9}
\, {\rm{Hz}}$, which leads to a difference in accumulated GW cycles of
$0.03$ rads after a one year evolution.

Finite mass-ratio corrections can be incorporated into the EOB model by
including subleading terms of ${\cal{O}}(m_{2}/m_{1})$ and
${\cal{O}}(q_{2})$ in the Hamiltonian, angular momentum and
$r(\omega)$ relation. We shall first ignore such terms to compare
against Teukolsky-based waveforms. In Sec.~\ref{sec:hi-order}, we
shall study how our results change when we include such terms. To do so, we
shall still assume circular, equatorial orbits and an adiabatic
evolution, but employ the spin EOB Hamiltonian of
Refs.~\cite{Barausse:2009aa,Barausse:2009xi} (reviewed in
Appendix~\ref{app:seob}), instead of the Kerr Hamiltonian of
Eq.~\eqref{HKerr}.

Except for this change, the EOB waveform modeling with finite mass-ratio 
corrections follows closely the derivation presented above. First, we
compute the angular momentum associated with the Hamiltonian of
Eq.~(\ref{hreal}) for circular, equatorial orbits, imposing $
\dot{p_r} = (\partial H_{\rm EOB}/\partial r)_{p_r=0}=0$ and solving
for $L \equiv p_\phi$. Then, we derive the orbital frequency $\omega =
(\partial H_{\rm EOB}/\partial p_\phi)_{p_r=0}$ to relate $r$ to
$\omega$, and to express $L$ in terms of $\omega$. When mass-ratio
corrections are present, however, the Hamiltonian becomes much more
involved, so solutions for $L$ as a function of $r$ must be searched
numerically. We have checked that the discretization and interpolation
used to solve these equations numerically do not introduce an error
larger than $10^{-10}$ in the Hamiltonian (\ref{hreal}).

%-----------------------------------------------------------------------------

\subsection{Numerical modeling: Teukolsky-based waveforms}
\label{sec:Nmodeling}

Teukolsky-based waveforms is the name we give to radiative models that use the Teukolsky equation to prescribe the radiation-reaction force for an inspiral. We use BH perturbation theory, considering a background spacetime with mass $m_1$ and spin $|{\bf S}_1| = m_1^{2} q$ (recall that the spin parameter $q_1 = a_1/m_1)$. The inspiraling object is a test-body with mass $m_{2} \ll m_{1}$ 
and no spin ($q_{2} = 0$). In principle, the masses and spins used here should be the same as those introduced 
in the EOB model. 

The radiative approximation assumes that EMRIs can be modeled as an adiabatic sequence of geodesics with slowly-varying {\it constants} of the motion. Consider the discretization of the orbital phase space, each point of which represents a certain geodesic with a given set of constants of motion (energy $E$,  angular momentum $L$ and Carter constant $Q$). For quasi-circular, equatorial EMRIs, the Carter constant vanishes, while the variation of the energy and angular momentum are related via $\delta L = (\delta E)/\omega$, $\omega$ being the orbital frequency. At each point in the orbital phase space, the geodesic equations can be solved to obtain the orbital trajectory of the small compact object, given any spin of the background.  

Once the geodesic trajectories are known, one can use these to solve the linearized Einstein equations and obtain the gravitational metric perturbation. This is best accomplished by rewriting the linearized Einstein equations in terms of the Newman-Penrose curvature scalar $\psi_{4}$ to yield the Teukolsky equation~\cite{Teukolsky:1973ap}. One can decompose $\psi_{4}$ into spin-weight $-2$ spheroidal harmonics ${}_{-2}S_{\ell m}^{a \omega}(\theta)$, using Boyer-Lindquist coordinates $(t,r,\theta,\phi)$, in the Fourier domain:
\begin{eqnarray}
\psi_{4} &=& \frac{1}{(r - i m_{1} q_{1} \cos{\theta})^{4}} \int_{-\infty}^{\infty} d\omega 
\nonumber \\
&\times& \sum_{\ell m} R_{\ell m \omega}(r) \; _{-2}S_{\ell m}^{m_{1} q_{1}  \omega}(\theta) e^{-i (m \phi - \omega t)}\,.
\end{eqnarray}
The radial functions $R_{\ell m \omega}(r)$ satisfy the radial Teukolsky equation
\begin{equation}
\Delta^{2} \frac{d}{dr} \left( \frac{1}{\Delta} \frac{dR_{\ell m \omega}}{dr} \right) - V(r) R_{\ell m \omega} = - {\cal{T}}_{\ell m \omega}\,
\end{equation}
where $\Delta$ is given in Eq.~(\ref{Delta}) and the radial potential is
\begin{equation}
V(r) \equiv - \frac{K^{2} + 4 i (r - m_{1}) K}{\Delta} + 8 i \omega r + \lambda\,,
\end{equation}
with $K \equiv (r^{2} + m_{1}^{2} q_{1}^{2}) \omega - m m_{1} q_{1}$, $\lambda \equiv {\cal{E}}_{\ell m} - 2 m_{1} q_{1} m \omega + m_{1}^{2} q_{1}^{2} \omega$, and ${\cal{E}}_{\ell m}$ the spheroidal harmonic eigenvalue. The source function ${\cal{T}}_{\ell m \omega}$ is given explicitly in Eq.~$(4.26)$ of 
Ref.~\cite{Hughes:1999bq} and it depends on the stress-energy tensor for a test-particle in a geodesic trajectory. 

The Teukolsky equation admits two asymptotic solutions: one outgoing as $r \to \infty$ and one ingoing as one approaches the background's event horizon. These two solutions represent outgoing radiation at future null infinity and ingoing radiation that falls into the BH through the event horizon. Both types of radiation are critical in the modeling of EMRIs; not including BH absorption can lead to errors in the waveform of order $10^{4}$ radians~\cite{Hughes:2000ssa,Hughes:2001jr}. These solutions can then be used to reconstruct both the GW radiated out to infinity, as well as the total energy flux lost in GWs. The energy flux can then be related to the temporal rate of change of the orbital elements, such as the orbital radius.

Solving the Teukolsky equation for a geodesic orbit tells us how that orbit tends to evolve due to the dissipative action of GW emission. By doing so for each point in orbital phase space, we endow this space with a set of vectors that indicate how the binary flows from one orbit to another.  We compute these vectors at a large number of points, and use cubic spline interpolation to estimate the rates of change of orbital constants between these points.  This allows us to compute the temporal evolution of all relevant quantities, including the orbital trajectories and gravitational waveforms.

We implemented this algorithm, discretizing the orbital phase space from an initial separation of $r = 10,000 \, m_{1}$ to the Kerr ISCO
\begin{eqnarray}
\frac{r_{\rm ISCO}}{m_{1}} &=& 3 + Z_{2} \mp \sqrt{(3 - Z_{1}) (3 + Z_{1} + 2 Z_{2})}\,,
\nonumber\\
Z_{1} &=& 1 + \left(1 - q_{1}^2 \right)^{1/3} \left[ \left(1 + q_{1}\right)^{1/3} + \left(1 - q_{1}\right)^{1/3}\right]\,,
\nonumber \\
Z_{2} &=& \left(3 q_{1}^{2} + Z_{1}^{2} \right)^{1/2}\,,
\end{eqnarray}
in a $1,000$ point grid, equally spaced in 
\begin{equation}
v \equiv (m_1\,\omega)^{1/3}= \left[\frac{q_{1}  - r^{3/2}/m_{1}^{3/2}}{q_{1}^2 - r^3/m_{1}^3} \right]^{1/3}\,. 
\end{equation}
We cannot evolve inside the Kerr ISCO with such a frequency-domain code, as stable orbits do not exists in this regime. We cannot evolve inside the Kerr ISCO with a frequency-domain code, as stable orbits do not exist in this regime (and so such orbits do not have a well-defined frequency spectrum).  The code we use to construct our waves is based on~\cite{Hughes:1999bq,Hughes:2001jr,2006PhRvD..73b4027D}, updated to use the spectral methods introduced by Fujita and Tagoshi~\cite{2004PThPh.112..415F,2005PThPh.113.1165F}.  A detailed presentation of this code and its results is in preparation~\cite{WT-SAH}.

The dominant error in these Teukolsky-based waveforms is due to truncation of the sums over $\ell$ and $m$. In particular, to compare with PN results, we must map this spheroidal decomposition to a spherical one (see Appendix~\ref{app:sph} for more details). Such a mapping requires one to include a buffer region of $\ell$ modes about the largest mode computed. We have been careful to use a sufficiently wide buffer and total number of $\ell$ modes such that the energy fluxes are accurate to $10^{-10}$ for all velocities and spins. In particular, this means that in the strong field region (close to the ISCO) up to $50 \, \ell$ modes were included. Other sources of error are due to the intrinsic double precision in the numerical solution to the Teukolsky equation, the discretization of the orbital phase space, and its cubic-spline interpolation. All of these amount to errors of order $10^{-14}$. All in all and in terms of GW phase, the Teukolsky-based waveforms are accurate to at least $10^{-2}$ rads.~during an entire year of evolution.

%-----------------------------------------------------------------------------
\section{Calibrating the test-particle energy flux}
\label{sec:calibration}
      
We consider here a calibrated EOB model that is built from the $h_{\ell m}$ functions in Eq.~\eqref{full-h}, 
but in which higher-order PN terms are included in the functions $\rho_{\ell m}$ and 
are calibrated to the numerical results. In particular, we write
\ba
\rho^{22}_{\rm Cal} &=& \rho^{22} +  \left[a_{22}^{(9,1)} + b_{22}^{(9,1)} \; 
{\rm{eulerlog}}_{2}v^{2}\right]\,\bar{q} \, v^9 
\nonumber \\
&+& \left[a_{22}^{(12,0)} + b_{22}^{(12,0)} \; {\rm{eulerlog}}_{2}v^{2} \right] v^{12}\,,
\nonumber \\
\rho^{33}_{\rm Cal} &=& \rho^{33} +  \left[a_{33}^{(8,2)} + b_{33}^{(8,2)} \; {\rm{eulerlog}}_{3}v^{2} \right] 
\,\bar{q}^2\, v^8 
\nonumber \\
&+& \left[a_{33}^{(10,0)} + b_{33}^{(10,0)} \; {\rm{eulerlog}}_{3}v^{2} \right] v^{10}\,,
\label{cal-of-f22}
\ea
where $(a_{\ell m}^{(N,M)},b_{\ell m}^{(N,M)})$ are
eulerlog-independent and eulerlog-dependent calibration coefficients
that enter the $(\ell,m)$ mode at ${\cal{O}}(v^{N})$ and proportional
to $\bar{q}^{M}$. As in Refs.~\cite{Damour:2008gu, Pan:2010hz} the
euler-log function is defined as
\begin{equation}
{\rm eulerlog}_m(x) = \gamma_E + \log\left(2m\sqrt{x}\right)\;,
\end{equation}
where
% ${\rm{eulerlog}}_{m}(x) \equiv
%\gamma_{E} + \log 2 + \log m + 1/2 \; \log x$, where
$\gamma_{E}=0.577215\ldots$ is Euler's constant. Notice that we have
introduced $4$ calibration parameters in the non-spinning sector of
the flux and $4$ in the spinning sector. The spin parameter $\bar{q}$
denotes here the spin of the background. When we neglect mass-ratio terms, we 
choose $\bar{q} = q_{1}$. However, when we switch on the mass-ratio terms we have an ambiguity on the 
choice of $\bar{q}$. Following Ref.~\cite{Pan:2010hz} we choose  $\bar{q} = q$, 
where $q$ is the deformed-Kerr spin parameter defined in Appendix ~\ref{app:seob}. Note that 
since $q =q_{1} + {\cal{O}}(m_{2}/m_{1})$, these choices are identical in the
test-particle limit, which is when the energy flux is calibrated. 

The choice of calibrating function in Eq.~\eqref{cal-of-f22} is rather
special and requires further discussion. We have chosen this function
so that leading-order corrections in the two dominant GW modes,
$(2,2)$ and $(3,3)$, are included. Higher $(\ell,m)$ modes contribute
significantly less to the GW and its associated flux. In each mode, we
have included the leading-order unknown terms that are both
$q$-independent and $q$-dependent. Since $q$-independent terms in
$\rho_{\ell m}$ are known to much higher order ($5.5$PN) than the
$q$-dependent ones ($4$PN) in the test-particle
limit~\cite{Shibata:1994jx,Tanaka:1997dj,Mino:1997bx},
spin-independent calibration coefficients enter at a much higher PN
order. The spin-dependence of the calibration terms is inferred from
known terms at lower PN orders. We have investigated many functional
forms for the calibrating functions, with a varying number of degrees
of freedom, and found the one above to be optimal in the class
studied.

The introduction of $8$ additional calibration terms might seem like a
lot. In Ref.~\cite{Yunes:2009ef}, the knowledge of non-spinning terms
up to $5.5$ PN order was found to be crucial to obtain a sufficiently
good agreement in the flux. Moreover, only $4$ additional calibration terms
($2$ at 6PN order in $\rho_{22}$ and $2$ at 5PN order in $\rho_{33}$)
were needed to reach a phase agreement of 1 rad after two years of
evolution. Similarly here, we expect that if the remaining $4.5$, $5$,
and $5.5$ PN order terms were calculated in the test-particle limit
when the central black carries a spin, then the flux would also
improve, requiring a smaller number of calibration parameters. It is
quite likely that those higher-order PN terms will be computed in the
near future, as they involve dramatically less complicated
calculations than PN terms in the comparable-mass case.  

Having in hand an improved GW energy flux carried away to infinity,
this must be enhanced with expressions for the GW energy flux that is
absorbed by the background BH. We do so here by simply adding the
Taylor-expanded form of the latter (see Appendix~\ref{app:flux-abs}) to
the flux of Eq.~\eqref{pd-flux}. The BH absorption terms in the GW
energy flux depend on polygamma functions, which are computationally
expensive to evaluate. We have empirically found that expanding this
function in $q\ll1$ to $30$th order is a sufficiently good
approximation for our purposes. When performing computationally
expensive calculations (like the fits described below) we shall employ
such expansions, but when solving for the orbital phase and when
computing the waveforms we shall return to the full polygamma
expressions.

The resulting EOB energy flux, including BH absorption terms, is then
calibrated via a two-dimensional, least-squares minimization relative
to numerical data obtained from Teukolsky-based calculations. The
fitting routine is two dimensional because when considering EMRIs
about spinning backgrounds, the flux depends on two independent
variables: the orbital velocity (or frequency or separation) and the
spin of the background. This, in turn, increases the number of points
that need to be used by more than an order of magnitude to properly 
calibrate Eq.~\eqref{cal-of-f22}. In all fits, we have assumed a data
variance of $10^{-11}$ for all velocities and spins and we have
required a relative accuracy of one part in $10^{8}$. Since the data
is now two-dimensional, one must search for a {\emph{global}} minimum
in $(q,v)$ space. After doing so, we find the calibration parameters
\begin{eqnarray}
a_{22}^{(9,1)} &=& -3.1092\,, \quad
b_{22}^{(9,1)} = -18.786\,, 
\\
a_{22}^{(12,0)} &=& 493.08\,, \quad
b_{22}^{(12,0)} = -247.89\,, 
\\
a_{33}^{(8,2)} &=& -17.310\,, \quad
b_{33}^{(8,2)} = 22.500\,, 
\\
a_{33}^{(10,0)} &=& -113.01\,, \quad
b_{33}^{(10,0)} = 28.125\,,
\end{eqnarray}

The computational cost of the calibrations performed in this paper is
much larger than those carried out in Ref.~\cite{Yunes:2009ef} for
the following reasons. First, we consider twice as many calibration
parameters as in Ref.~\cite{Yunes:2009ef}, increasing the
dimensionality of the fitting space. Second, global minimization
routines require non-trivial algorithms that are numerically more
expensive than those employed in one-dimensional minimizations. Third,
the amount of data fitted increases by at least one order of
magnitude, due to the intrinsic bi-dimensionality of the
problem. Combining all of this, the computational cost of performing
the calibration is now more than $100$ times larger than in
Ref.~\cite{Yunes:2009ef}. Even then, however, these fits require
${\cal{O}}(10)$ CPU minutes to complete. Once they have been carried
out, this calculation does not need to be repeated again in the
waveform modeling.

\begin{figure}
\includegraphics[width=8.5cm,clip=true]{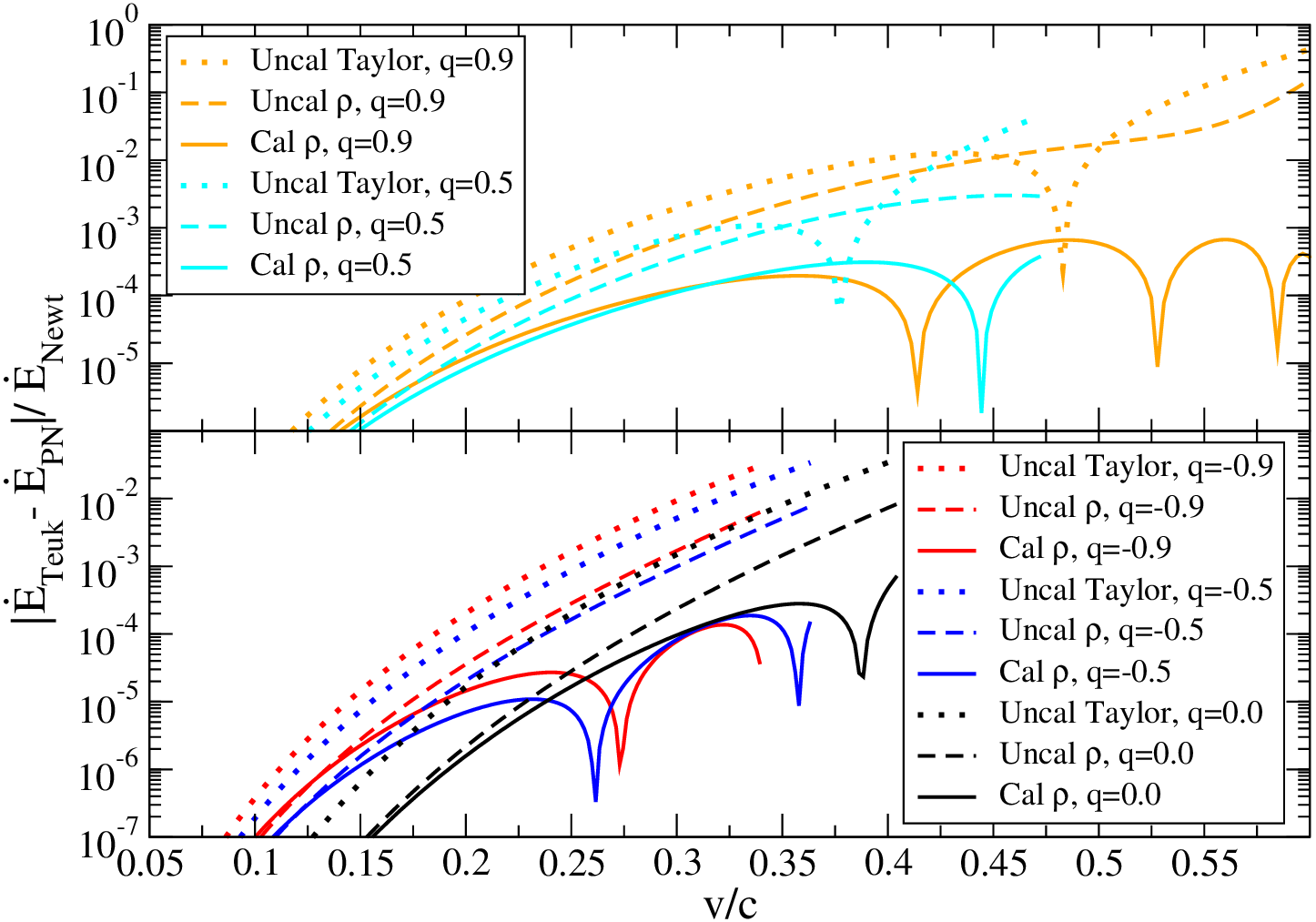} 
 \caption{\label{fig:flux-comp} Fractional difference between PN and Teukolsky-based fluxes as a function of velocity for spins $q = (0.5,0.9)$ (top) and $q = (-0.9,-0.5,0.0)$ (bottom). The dotted curves employ the Taylor-expanded PN flux with BH absorption terms, while the dashed and solid curves use the uncalibrated and calibrated $\rho$-resummed fluxes with BH absorption terms respectively.}
\end{figure}

Figure~\ref{fig:flux-comp} plots the fractional difference between the
analytical GW energy flux and that computed with Teukolsky-based waveforms as
a function of velocity, from an initial value of $v/c = 0.01$ to the
velocity at the ISCO, for five different spin values: $q =
(-0.9,-0.5,0.0,0.5,0.9)$. All comparisons are here normalized to the
Newtonian value of the flux $F_{\rm Newt} = 32/5 \nu^{2} v^{10}$.
The different curve styles differentiate between analytical
models: the dotted curves use the total, uncalibrated
Taylor-expansion; the dashed curves use the uncalibrated
$\rho$-resummed flux with BH absorption terms; the solid curves use
the calibrated $\rho$-resummed flux with BH absorption terms. Notice
that the calibrated model does better than the other two by at least
two orders of magnitude near the ISCO for all spin-values.

Several interesting conclusions can be drawn from
Fig.~\ref{fig:flux-comp}. First, as obtained in Ref.~\cite{Pan:2010hz}
the uncalibrated $\rho$-resummed model is better than the
Taylor-expanded version of the flux, by up to nearly an order of
magnitude at the ISCO for all spins. In turn, the calibrated model is
better than the uncalibrated one by one to two orders of magnitude
near ISCO for all spins. One could also calibrate the Taylor-expanded
flux (not shown in Fig.~\ref{fig:flux-comp}), but this would not
produce such good agreement in the entire $(v,q)$ space. This is
clearly because the uncalibrated $\rho$-resummed model is more
accurate than the Taylor one, and thus the calibration terms have to
do less work to improve the agreement. For the calibrated Taylor and
$\rho$-resummed models to become comparable in accuracy one would have
to include up to at least $16$ calibration coefficients in the Taylor
model.

\begin{figure}
 \includegraphics[width=8.5cm,clip=true]{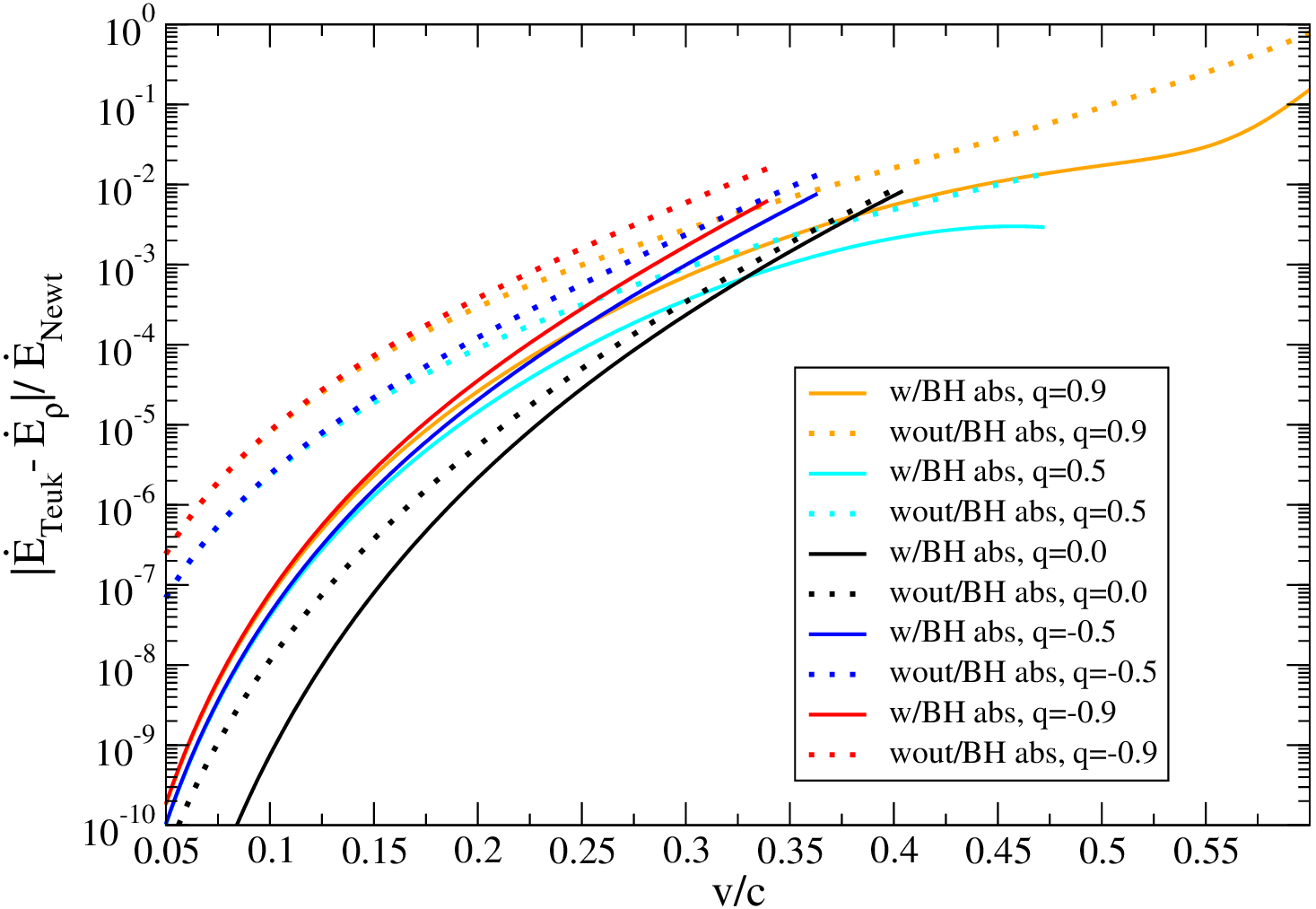} 
 \caption{\label{fig:BHAbs} Fractional difference between $\rho$-resummed and Teukolsky-based fluxes as a function of velocity for spins $q = (-0.9,-0.5,0.0,0.5,0.9)$. The dotted curves do not include the Taylor-expanded BH absorption contributions, while the solid lines do.}
\end{figure}

The inclusion of BH absorption coefficients is crucial to obtain good
agreement with the full Teukolsky-based flux, a result that was not
obvious for the case of non-spinning EMRIs.  Figure~\ref{fig:BHAbs}
plots the fractional difference between the uncalibrated EOB GW energy
flux and Teukolsky-based one as a function of velocity for five different 
spin values: $q = (-0.9,-0.5,0.0,0.5,0.9)$, from an initial 
value of $v/c = 0.01$ at the ISCO, for five different spin values: 
$q = (-0.9, -0.5, 0.0, 0.5, 0.9)$.  For these cases, we have 
$v_{\rm ISCO} = (0.343,0.367,0.408,0.477,0.609)$.
The solid curves use the uncalibrated EOB model including the
Taylor-expanded BH absorption contributions, while the dotted curves
do not. For the non-spinning case, observe that there is a very small
difference (smaller than $10^{-2}$) between adding the BH absorption
terms or not. 

For the spinning cases, however, this is not the
case. For rapidly spinning backgrounds, adding the BH absorption terms
improves the agreement by an order of magnitude. Presumably, resumming
these terms in a multipolar-factorized manner would improve the
agreement even more. The BH absorption terms play a much larger role in the spinning case because spin changes the PN order at which absorption enters in the energy flux.  These terms enter at 4PN order for Schwarzschild black holes, but at 2.5PN order for non-zero spin.  This change of order has a very large and important impact on the system's evolution.

The inclusion of calibration parameters to improve the agreement of
PN-inspired fluxes and Teukolsky-based ones for EMRIs is certainly not
new. In Ref.~\cite{Gair:2005ih}, a similar, PN-inspired calibration
was carried out for circular-inclined orbits (and more generic
ones). Before calibration, their fluxes were Taylor-expanded to 2PN order
and included only the contribution that escapes to infinity (not the BH absorption
terms discussed above). Their fit was then done with Teukolsky-data produced by an 
older version of the code used here, which was accurate only to one part in
$10^{6}$. Moreover, the fit was done in the range $r \in (5,30) \, M$
[$v \in (0.183,0.436)$], so the fitted function loses accuracy
rapidly outside this regime, particularly close to the ISCO. Inside the fitting regime,
the flux was fitted to an accuracy of $3 \times 10^{-2}$ using $45$ calibration
coefficients for an inclined, but fixed orbit. The accuracy decreases 
to $0.1$ for orbits which get closer to the ISCO. 

To fairly compare the results of Ref.~\cite{Gair:2005ih} with 
our results which are restricted to circular, equatorial orbits, 
we implemented their model and re-calibrated it considering 
only circular, equatorial orbits. Using $45$ coefficients, 
we found an accuracy similar to ours at high velocities close to the ISCO, 
but worse at low velocities. This is because their fluxes before calibration 
are not as accurate as the one employed here (by including up to $5.5$PN 
order terms and BH absorption terms), particularly at low velocities. It is important
to emphasize that by calibrating $8$ parameters instead of $45$ we here obtain
better flux accuracies than in Ref.~\cite{Gair:2005ih} for circular, equatorial EMRIs. 
We could obtain even better accuracy if we were using a larger number of calibration 
coefficients, e.g.~using $16$ coefficients the agreement with the Teukolsky-based 
flux would be of ${\cal{O}}(10^{-5})$.

%-----------------------------------------------------------------------------
\section{Comparison of the GW phase and amplitude}
\label{sec:results}

The comparison of EOB and Teukolsky evolutions requires that we choose
a specific EMRI. We shall here follow Ref.~\cite{Yunes:2009ef} and
choose system parameters that define two classes of EMRIs:
\begin{itemize}
\item {\bf{System I}} explores a region between orbital separations
  $r/M \in (16,26)$, which spans orbital velocities and GW frequencies
  in the range $v \in (0.2,0.25)$ and $f_{\rm GW} \in (0.005,0.01) \;
  {\rm{Hz}}$ respectively. Such an EMRI has masses $m_{1} = 10^{5} \,
  M_{\odot}$ and $m_{2} = 10 \, M_{\odot}$ for a mass ratio of
  $10^{-4}$ and it inspirals for $\sim (6.3\mbox{--}6.7) \times 10^{5}$ rads
  of orbital phase depending on the spin.
\item {\bf{System II}} explores a region between orbital separations
  $r/M \in (11,r_{\rm ISCO})$, which spans orbital velocities and GW
  frequencies in the range $v \in (0.3,v_{\rm ISCO})$ and $f_{\rm
    GW} \in (0.001,f_{\rm GW}^{\rm ISCO}) \; {\rm{Hz}}$
  respectively. Such an EMRI has masses $m_{1} = 10^{6} \, M_{\odot}$
  and $m_{2} = 10 \, M_{\odot}$ for a mass ratio of $10^{-5}$ and it
  inspirals for $\sim (1.9\mbox{--}4.5) \times 10^{5}$ rads of orbital phase
  depending on the spin.
\end{itemize}
The evolution of Sys.~I is stopped around an orbital separation of $16
M$, because this coincides with a GW frequency of $0.01 \; {\rm{Hz}}$,
which is close to the end of the LISA sensitivity band. The evolution
of Sys.~II is usually stopped at the orbital separation corresponding
to the ISCO, or whenever its GWs reach a frequency of $0.01 \;
{\rm{Hz}}$. For each of these systems, we shall investigate different
background spin parameters.

Before proceeding, notice that Sys.~I and II should not be compared on
a one-to-one basis. One might be tempted to do so, as Sys.~I resembles
a weak-field EMRI, which inspirals at a larger orbital separation and
with smaller orbital velocities than Sys.~II, a more strong-field
EMRI. Comparisons are not straightforward, however, as these systems
accumulate a different total number of GW cycles. In fact, Sys.~I
usually accumulates almost twice as many GW cycles as
Sys.~II. Therefore, even though one might expect PN models of Sys.~I
to agree better with Teukolsky-based evolutions, this need not be the
case, as this system has more {\it time} (as measured in GW cycles) to
accumulate a phase and amplitude difference than Sys.~II.

We compare the EOB and the Teukolsky-based waveforms after aligning
them in time and phase. Such an alignment is done by minimizing the
statistic in Eq.~(23) of Ref.~\cite{Buonanno:2009qa}, just as was done
in Ref.~\cite{Yunes:2009ef}. This is equivalent to maximizing the
fitting factor over time and phase of coalescence in a matched
filtering calculation with white noise~\cite{Buonanno:2009qa}. The
alignment is done in the low-frequency regime, inside the time
interval $(0,64) \lambda_{\GW}$, where $\lambda_{\GW}$ is the GW
wavelength. This quantity depends on the spin of the background,
ranging from $386 \, M$ ($63 \, M$) to $415 \, M$ ($121 \, M$) for
Sys.~I (Sys.~II). This corresponds to aligning the initial phase and
frequency inside a window of length in the range $(0.004,0.01)$ months
depending on the system and spin of the background. We have checked
that increasing the size of the alignment window does not affect the
final phase and amplitude difference; for example, for a spin of $q=
0.9$ and Sys.~I, increasing the alignment window by a factor of two
changes the final phase difference by $0.002$ rads and the relative,
fractional amplitude agreement by $0.0004 \%$.

\begin{figure*}
\begin{center}
\begin{tabular}{cc}
 \includegraphics[width=8.5cm,clip=true]{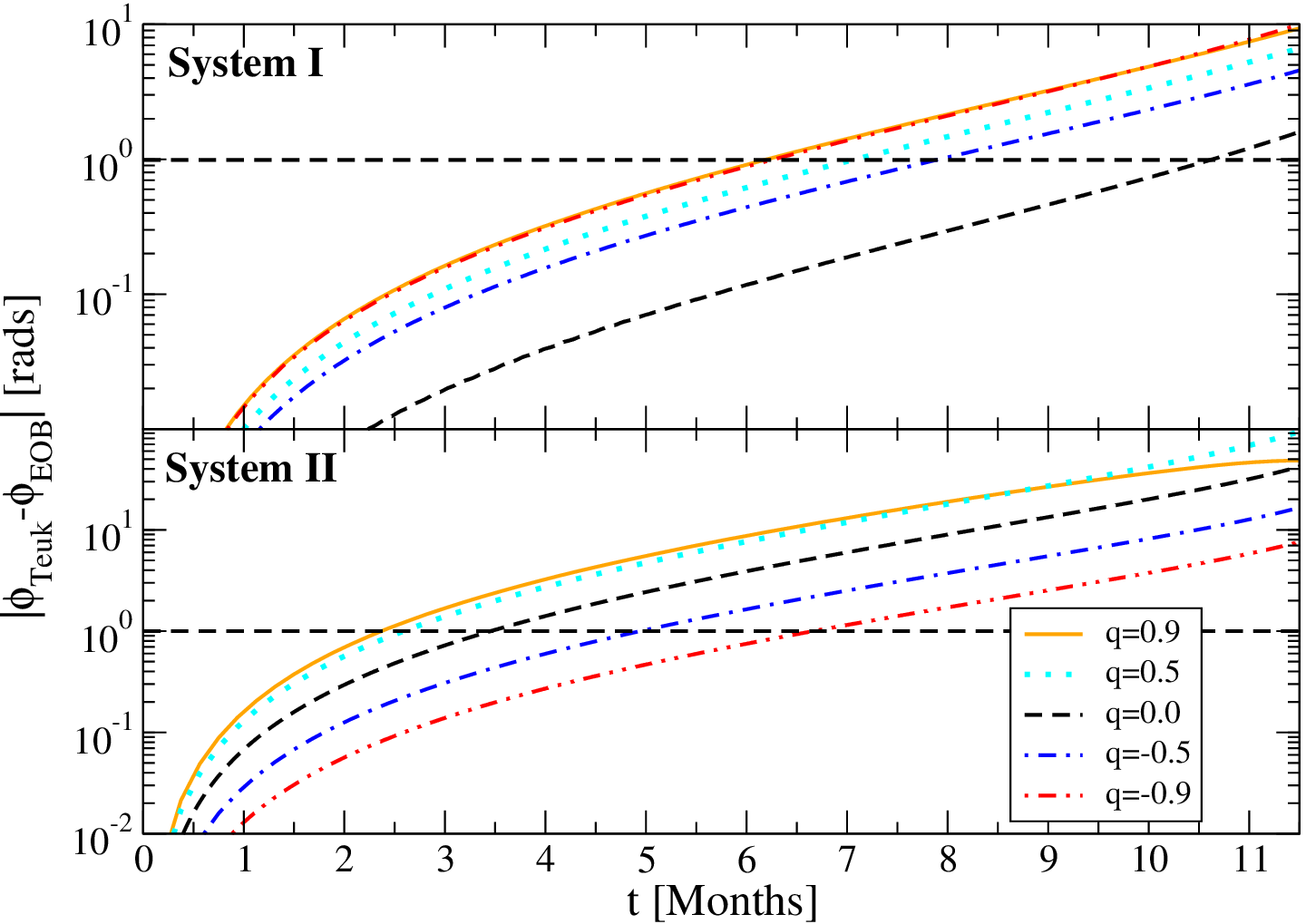} \quad 
  \includegraphics[width=8.5cm,clip=true]{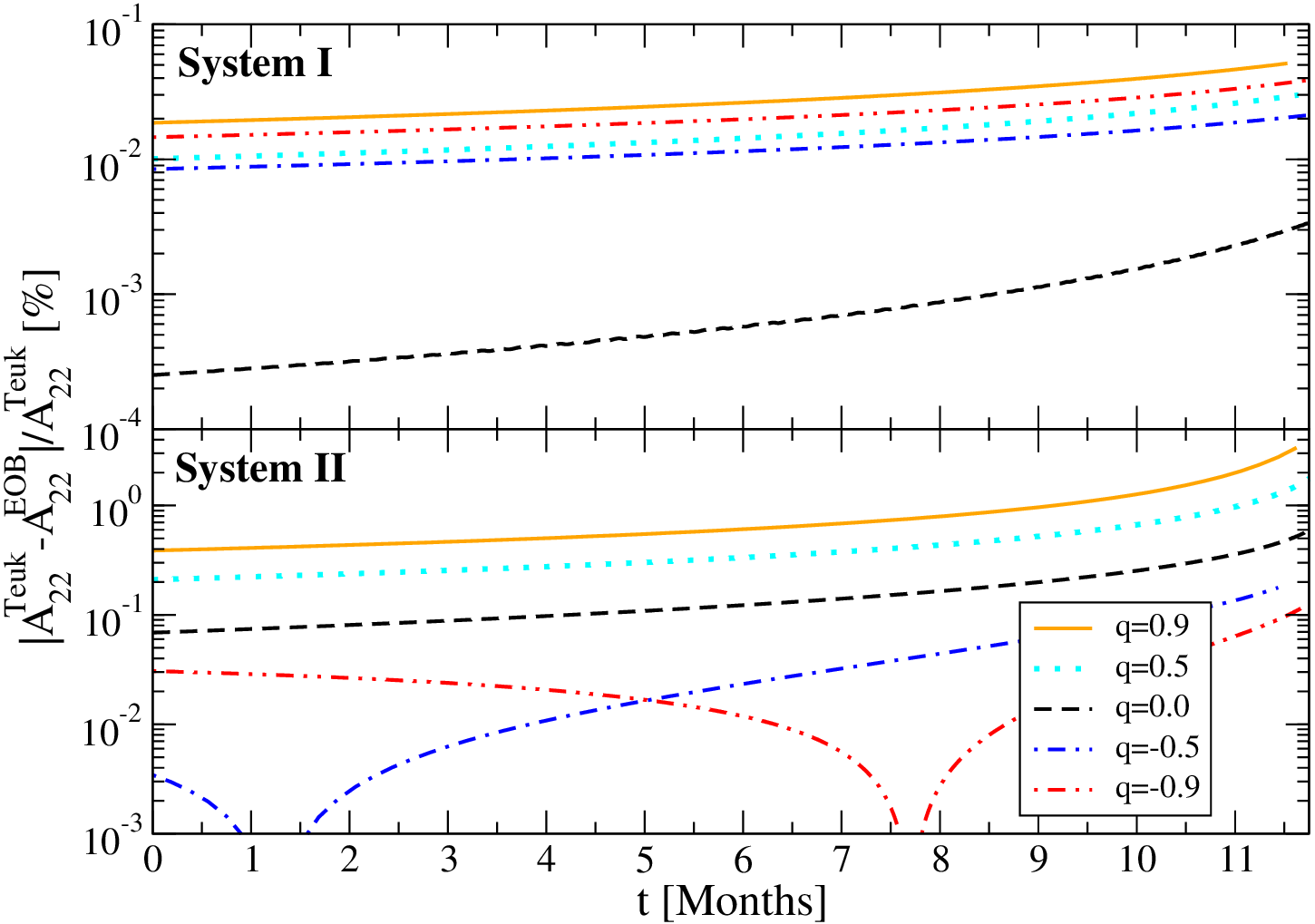}  
\end{tabular}
\end{center}
 \vspace{-0.4truecm}
 \caption{\label{fig:waveform-comp}  Absolute value of the dephasing (left) and relative, fractional amplitude difference (right) computed in the calibrated EOB model and the Teukolsky-based waveforms for the dominant $(\ell,m)=(2,2)$ mode. Different curves correspond to different background spin values. }
\vspace{-0.4truecm}
\end{figure*}
Figure~\ref{fig:waveform-comp} shows the absolute value of the
dephasing and relative, fractional amplitude difference of the
dominant $(\ell,m)=(2,2)$ mode for both systems and a variety of
background spins. For Sys.~I, the calibrated EOB model maintains a 1
radian phase accuracy over at least the first 6 months for all spin
values, while for Sys.~II the same phase accuracy is maintained for up
to only the first 2 months. The amplitude agreement is also excellent
for all spin values, with better agreement for Sys.~I. As found in
Ref.~\cite{Yunes:2009ef} the GW phase agreement is primarily due to
the correct modeling of the orbital phase, as the former tracks the
latter extremely closely; we find that the difference between the
orbital and GW phase over a one year evolution is less than $0.1$
rads.

\begin{figure}[h]
 \includegraphics[width=8.5cm,clip=true]{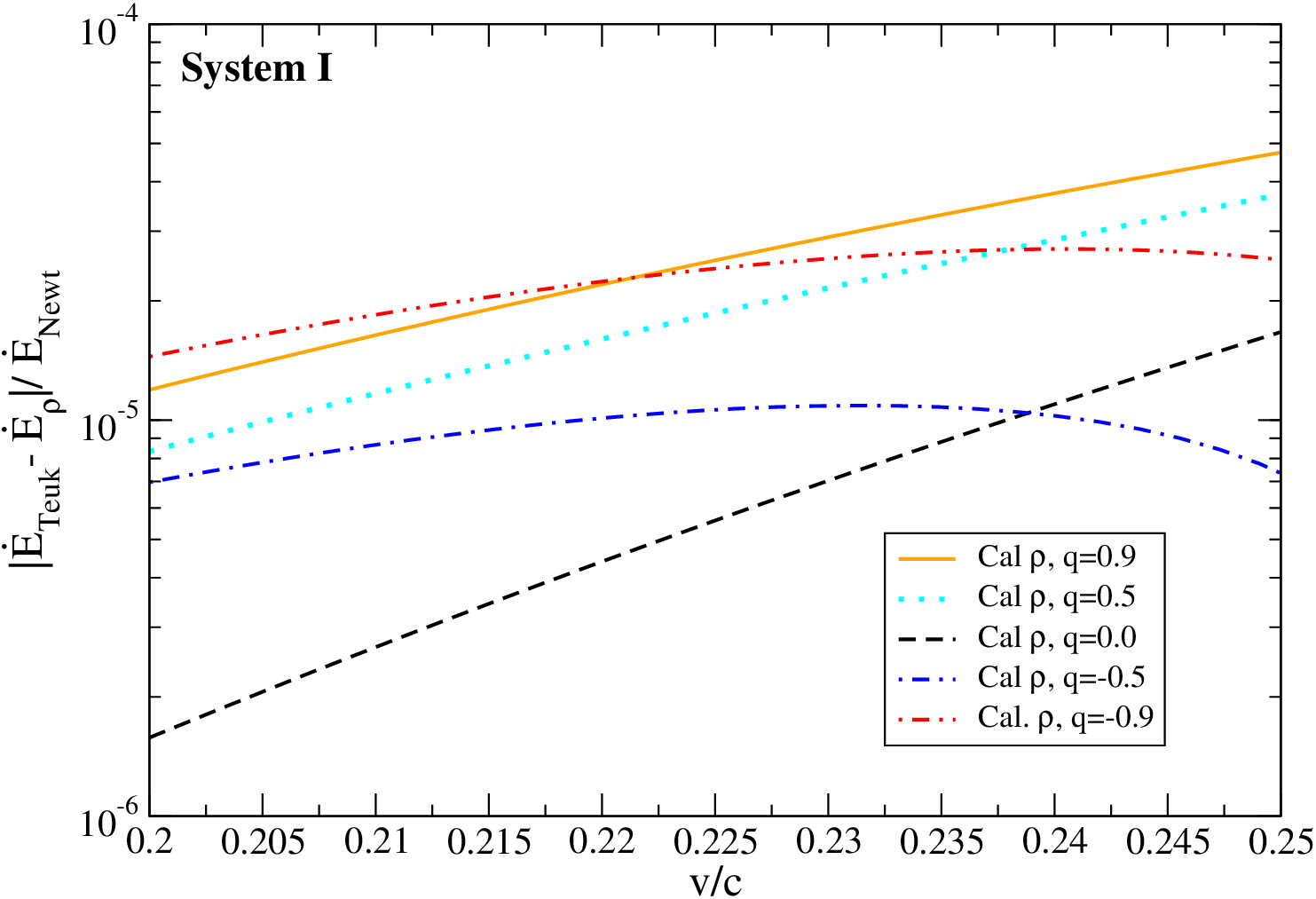} 
 \caption{\label{fig:flux-comp-zoom} Fractional difference between the calibrated $\rho$-resummed and Teukolsky-based fluxes for spins $q = (-0.9,-0.5,0.0,0.5,0.9)$ as a function of velocity. We plot here only the range of velocities sampled by Sys.~I.}
\end{figure}

The agreement in the phase as a function of background spin follows
closely the flux agreement shown in Fig.~\ref{fig:flux-comp}. This is
hard to see in Fig.~\ref{fig:flux-comp}, which is why
Fig.~\ref{fig:flux-comp-zoom} zooms into the velocity region sampled
by Sys.~I and plots all background spin cases for the calibrated
$\rho$-resummed system. Observe that the $q=0.0$ case has the best
flux agreement, which explains why the phase and amplitude agreement
is so good for this case in Fig.~\ref{fig:waveform-comp}. Observe also
that the $q=0.9$ and $q=-0.9$ cases have the worst flux agreement,
which also explains why they disagree the most in phase and amplitude
in Fig.~\ref{fig:waveform-comp}.

The accuracy of the calibrated EOB model is excellent relative to
Taylor-expanded PN models. If one were to use an uncalibrated
Taylor-expanded version of the flux, instead of the calibrated
$\rho$-resummed flux, one would find a phase and amplitude
disagreement of $\sim 10^{1}\mbox{--}10^{2}$ rads [$\sim
10^{3}\mbox{--}10^{4}$ rads] and $\sim 0.1 \%$ ($\sim 10 \%$) for
Sys.~I (Sys.~II) after a one year-evolution for different spin
values. The above results are consistent with the arguments in
Ref.~\cite{Mandel:2008bc}, who concluded that $3.5$PN accurate GW
phase expressions could lead to phase errors around $10^{3}$--$10^{4}$
radians over the last year of inspiral. That analysis reached those
conclusions by comparing $3$ to $3.5$PN accurate, analytic expressions
for the GW phase. Here, we are comparing full-numerical evolutions of
the PN equations of motion carried out to much higher order, and, of
course, we find that such conclusions depend sensitively on the type
of EMRI considered and the spin of the background.

\begin{figure}[h]
 \includegraphics[width=8.5cm,clip=true]{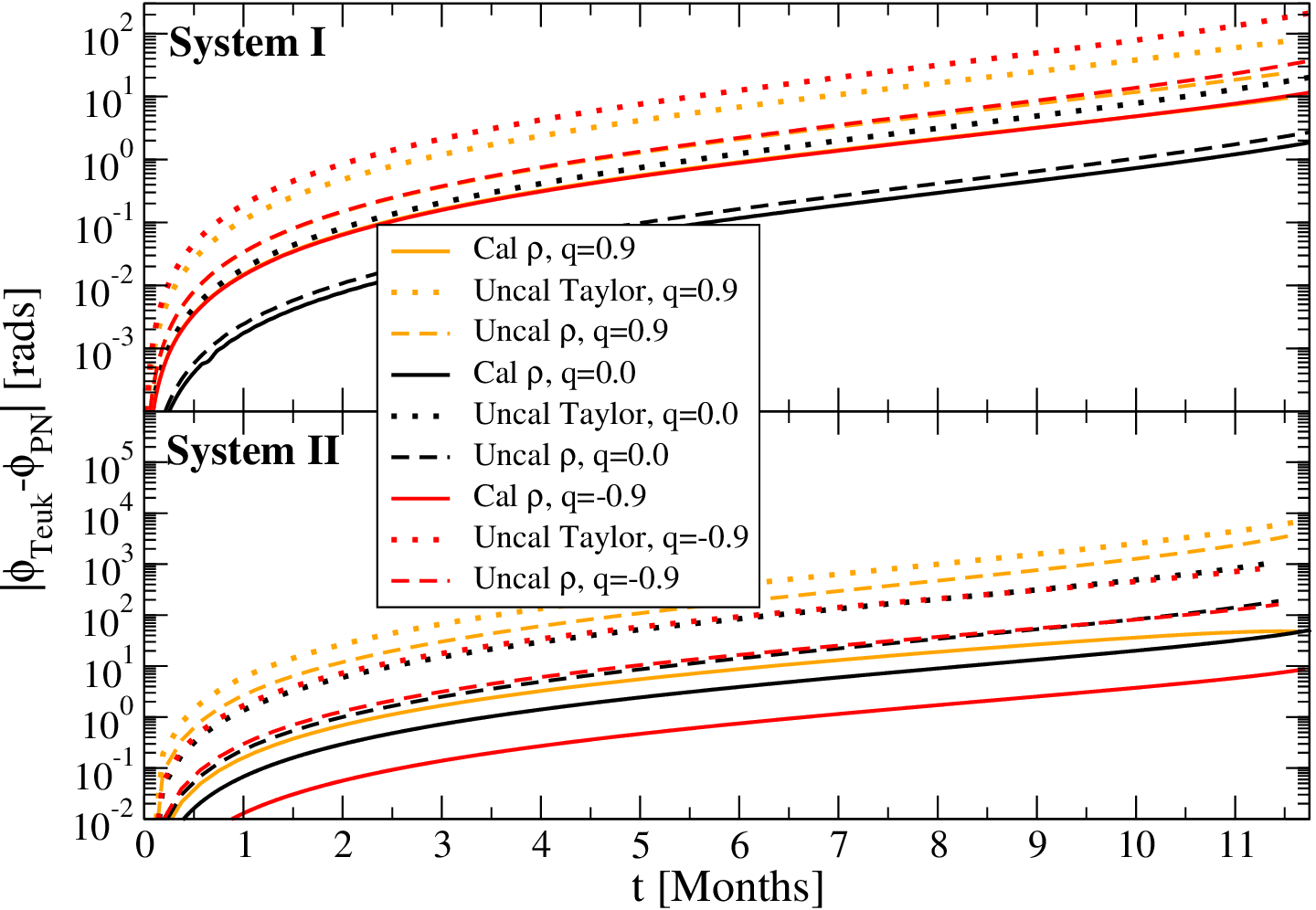} 
 \caption{\label{fig:waveform-comp-uncals}  Absolute value of the dephasing computed for the dominant mode in different PN models and the Teukolsky-based waveforms. Different curve colors/shades correspond to different background spin values, while different curve types correspond to different PN models.}
\end{figure}
The increase in accuracy of the calibrated $\rho$-resummed model is
due both to the calibration and to the $h_{\ell m}$ factorized
resummation. This fact can be appreciated in
Fig.~\ref{fig:waveform-comp-uncals}, where we plot the absolute value
of the dephasing for the dominant mode in different PN models and the
Teukolsky-based waveforms. Light curves (orange in the color version) correspond to background
spins of $q=0.9$, dark curves (red in the color version) to $q=-0.9$ and black curves to
non-spinning backgrounds. Dotted curves use the uncalibrated Taylor
flux model, dashed curves the uncalibrated $\rho$-resummed model and
solid curves the calibrated version. For Sys.~I, there is a large gain
in accuracy by switching from the uncalibrated Taylor model to the
uncalibrated $\rho$-resummed model, but then the calibration itself
does not appear to improve the accuracy substantially. For Sys.~II, on the
other hand, the calibration can increase the accuracy up to almost 2
orders of magnitude, as in the $q=0.9$ case.

\begin{figure*}
\begin{center}
\begin{tabular}{cc}
 \includegraphics[width=8.5cm,clip=true]{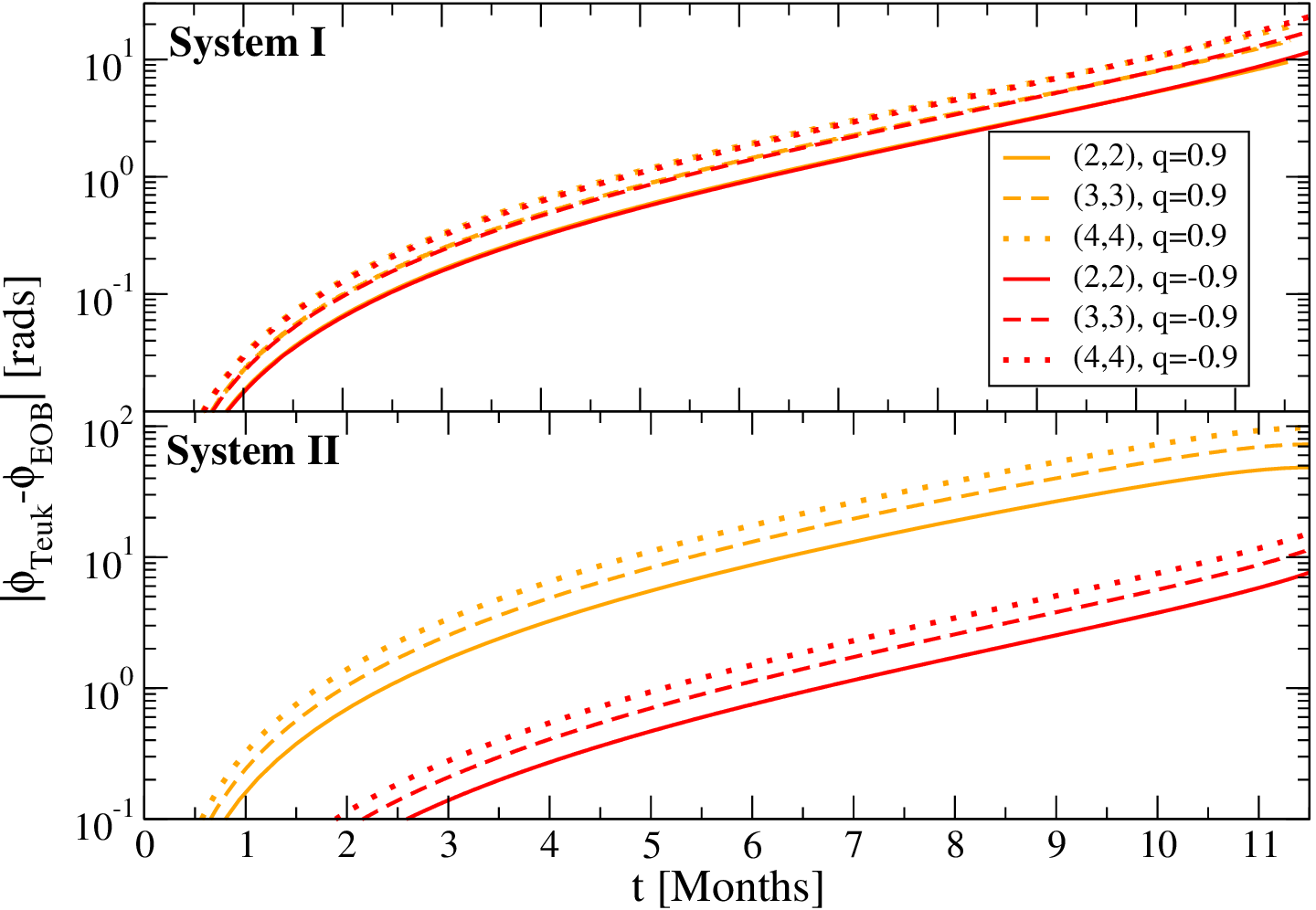} \quad 
  \includegraphics[width=8.5cm,clip=true]{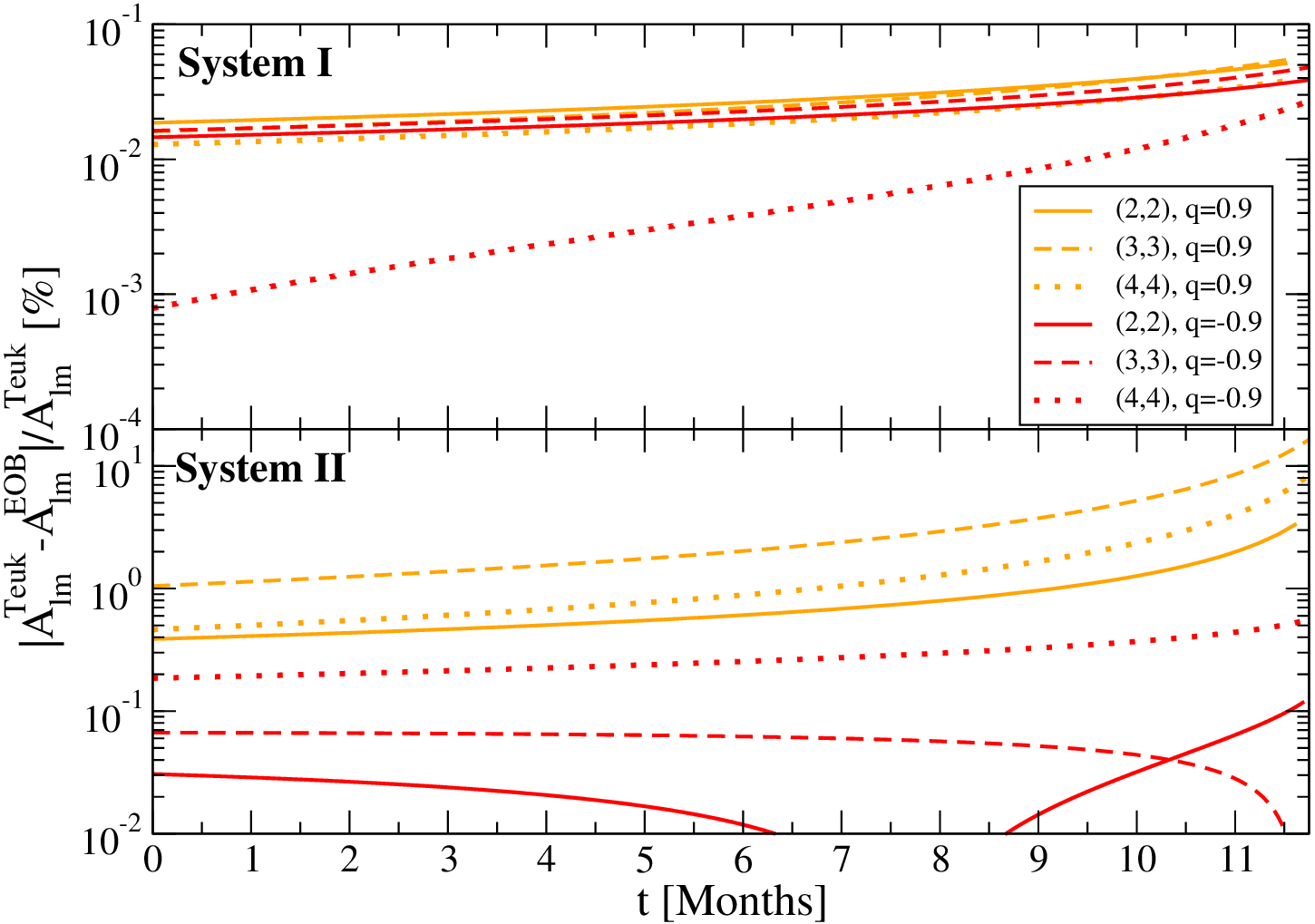}  
\end{tabular}
\end{center}
 \vspace{-0.4truecm}
 \caption{\label{fig:waveform-comp-higher-modes}  Absolute value of the dephasing (left) and relative, fractional amplitude difference (right) computed in the calibrated EOB model and the Teukolsky-based waveforms. The solid curve corresponds to the dominant $(2,2)$ mode, while the dashed curve is for the $(3,3)$ mode and the dotted curve for the $(4,4)$ mode. Different curves stand for different background spins.}
\vspace{-0.4truecm}
\end{figure*}

The agreement in the phase and amplitude is not only present in the
dominant mode, but also in higher $(\ell,m)$ ones, as shown in
Fig.~\ref{fig:waveform-comp-higher-modes}. We here plot the absolute
value of the dephasing and the relative, fractional amplitude
difference between the calibrated EOB model and Teukolsky-based
waveforms for the dominant mode, as well as the $(3,3)$ and $(4,4)$
modes. We have here shifted the higher $(\ell,m)$ modes using the best 
frequency shift that maximizes the agreement for the dominant mode. 
This agreement is simply a manifestation of the agreement in
the orbital phase. In fact, we find that $\Phi_{\rm GW} \sim m \Phi_{\rm orb}$,
with differences that are always less than $1$ rad for all systems considered. 

\begin{figure}
 \includegraphics[width=8.5cm,clip=true]{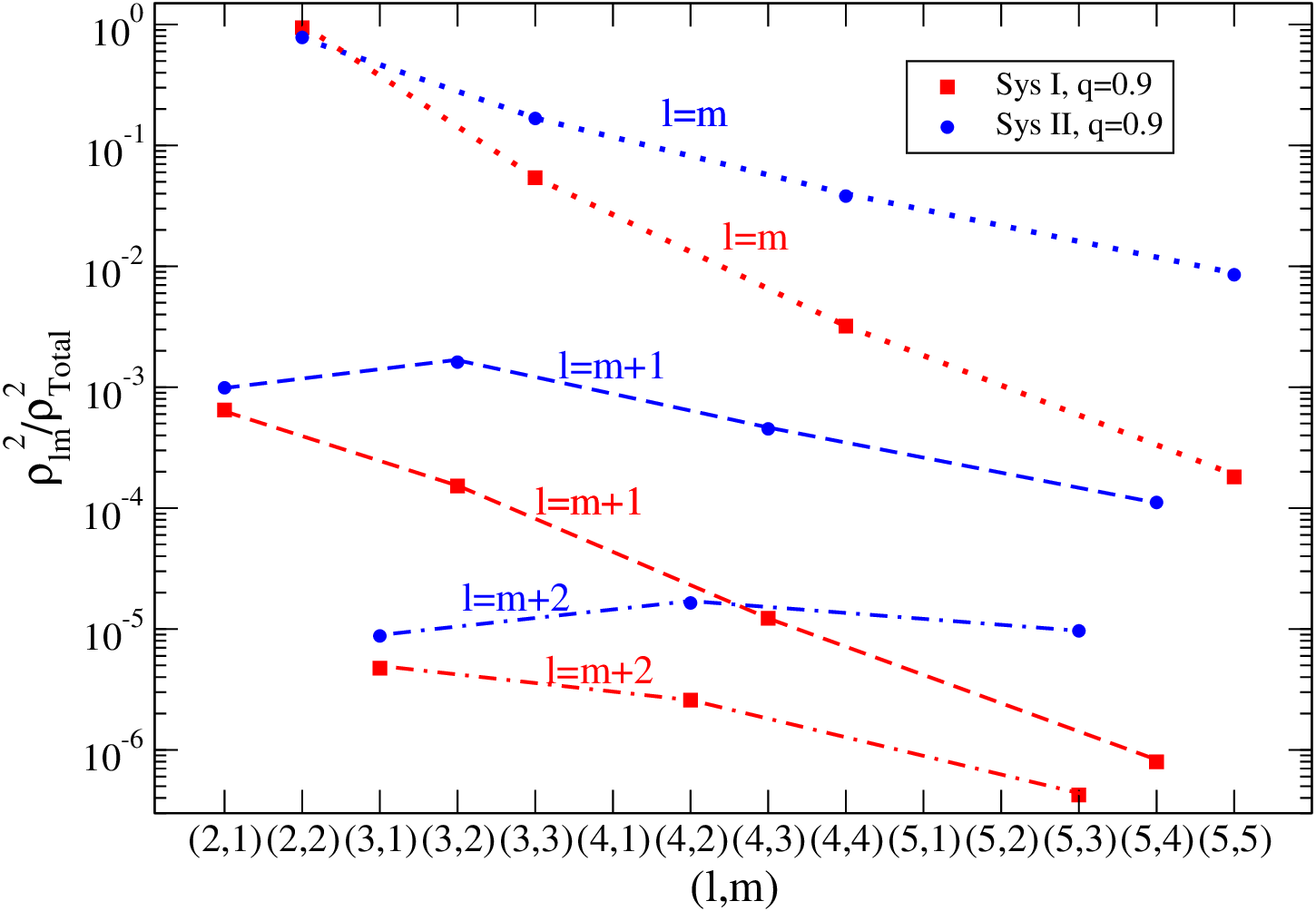}
 \caption{\label{fig:SNR} Relative fraction between the squared of the SNR computed with only the 
 $(\ell, m)$ mode and that computed with all $(\ell, m)$ modes. The lines connecting points are only meant to group points with $\ell = m$ (dotted), $\ell = m+1$ (dashed) and $\ell = m+2$ (dot-dashed).}
\end{figure}

Higher-$\ell$ modes contribute significantly less to the SNR than the dominant $(2,2)$ mode. 
Figure~\ref{fig:SNR} plots the relative fraction between the squared of the 
SNR computed with only the $h_{\ell m}$ component of the waveform and that computed 
by summing over all modes. This figure uses data corresponding to Systems I and II, both with
spin $q = 0.9$ (results for other spin values are almost identical). Clearly, the $(2,2)$ mode is 
dominant, followed by the $(3,3)$ and $(4,4)$ modes. Because of this feature of quasi-circular 
inspirals, obtaining agreement for the $(2,2)$ mode implies one can recover over $97 \%$ 
of the SNR. 
%

%-----------------------------------------------------------------------------
\section{Data Analysis Implications}
\label{sec:DA}

Although the phase agreement presented in the previous section is a good indicator of the validity of the EOB
model, one is really interested in computing more realistic data analysis measures. In this section we 
compute the mismatch between the Teukolsky and the EOB model, maximized over extrinsic parameters and as a function of observation time. 

Let us first introduce some basic terminology. Given any time series
$a(t)$ and $b(t)$, we can define the following inner-product in signal
space
\be
\left( a \right| \left. b \right) = 4\, {\rm Re} \int_{0}^{\infty} \frac{\tilde{a}(f) \; \tilde{b}^{\star}(f)}{S_{n}(f)}\,
\ee
where the overhead tildes stand for the Fourier transform and the star stands for complex conjugation. The quantity $S_{n}(f)$ is the spectral noise density curve, where here we follow~\cite{2004PhRvD..70l2002B,2005PhRvD..71h4025B}. Notice that we use the {\emph{sky-averaged}} version of this noise curve here, which is larger than the non-sky-averaged version by a factor of $20/3$. In particular, this means that our SNRs are smaller than those one would obtain with a non-sky-averaged noise curve by a factor of $(20/3)^{1/2} \sim 2.6$. 
%The SNRs computed here use the spectral noise density curve in~\cite{2004PhRvD..70l2002B}. Notice that this $S_{n}(f)$ differs from that of~\cite{BC} by a multiplicative factor of $20/3$ that arises due to the sky-averaging prescription. Our SNRs are thus larger than what one would find if using~\cite{BC} by a factor of $(20/3)^{1/2} \sim 2.6$

Given this inner-product, we can now define some useful measures. The SNR of signal $a$ is simply
\be
\rho = \sqrt{\left( a \right| \left. a \right)}\,,
\ee
while the overlap between signals $a$ and $b$ is simply
\be
{\rm M} = {\rm{max}} \frac{\left( a \right| \left. b \right)}{\sqrt{{\left( a \right| \left. a \right) {\left( b \right| \left. b \right)}}}}\,.
\ee
with the mismatch ${\rm MM} = 1 - {\rm M}$. The max label here is to remind us that this statistic must be maximized
over a time shift and a phase shift (see eg.~Appendix B of~\cite{Damour:1997ub} for a more detailed discussion). 

\begin{figure*}
\begin{center}
\begin{tabular}{cc}
 \includegraphics[width=8.5cm,clip=true]{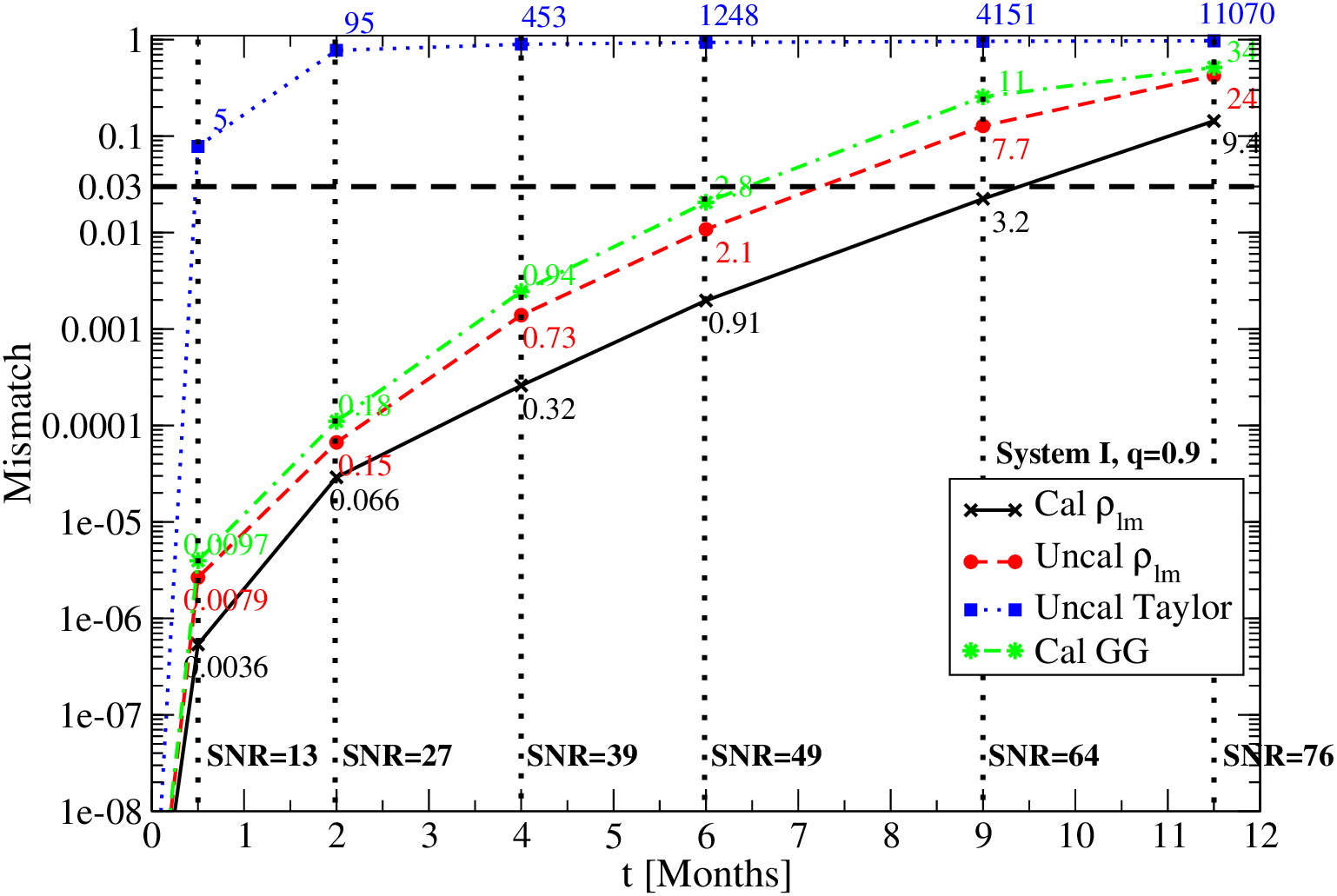} \quad 
  \includegraphics[width=8.5cm,clip=true]{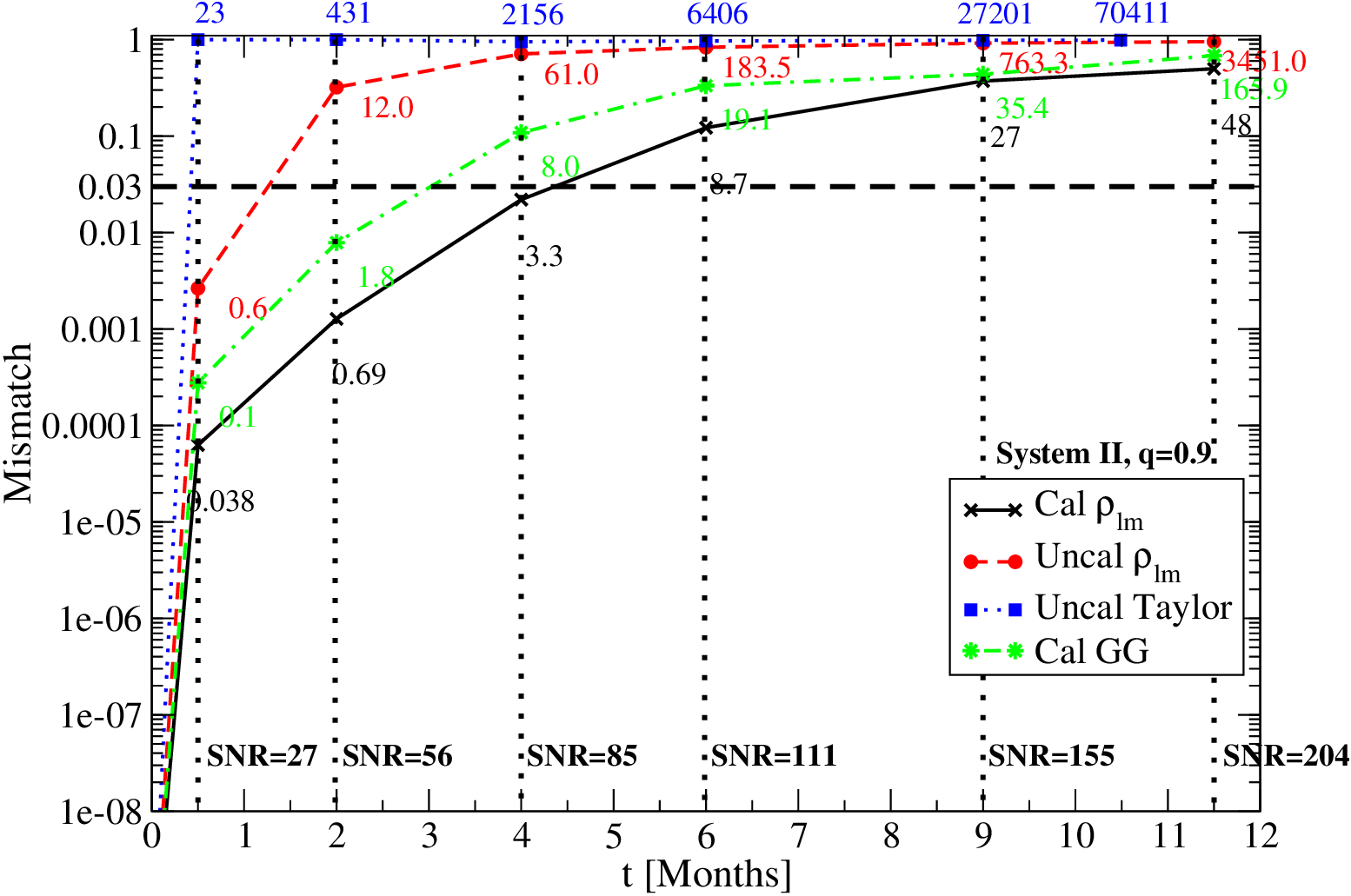}  
\end{tabular}
\end{center}
 \vspace{-0.4truecm}
 \caption{\label{fig:overlap} Mismatch between Teukolsky and EOB waveforms as a function of waveform duration for Systems I and II and a background spin of $q=0.9$.}
\vspace{-0.4truecm}
\end{figure*}

The data analysis measures introduced above ($\rho$ and ${\rm MM}$) depend
on the length of the time-series, i.e.~the observation
time. Figure~\ref{fig:overlap} plots the mismatch between the
Teukolsky-based waveforms and a variety of models for both Sys.~I and II and
a background spin of $q = 0.9$ as a function of observation time.  The
vertical lines correspond to observation times of $2$ weeks, $2$
months, $6$ months, $9$ months and $11.5$ moths, together with their
associated SNRs at 1 Gpc.
%~\footnote{The SNRs computed here use the spectral noise density curve in~\cite{2004PhRvD..70l2002B}. Notice that this $S_{n}(f)$ differs from that of~\cite{BC} by a multiplicative factor of $20/3$ that arises due to the sky-averaging prescription. Our SNRs are thus larger than what one would find if using~\cite{BC} by a factor of $(20/3)^{1/2} \sim 2.6$.}
The mismatches are computed with different
analytical models: black crosses stand for the calibrated $\rho$-model 
with $8$ calibration coefficients; red circles to the uncalibrated $\rho$- model; blue
squares to the uncalibrated Taylor model; green circles to that EOB
evolution using the original flux of Ref.~\cite{Gair:2005ih} which has $45$ calibration coefficients (denoted $GG$ in the figure). 
For comparison, we also include the amount of dephasing (numbers next to data points in Fig.~\ref{fig:overlap}) at $2$ weeks, $2$ months, $6$ months, 
$9$ months and $11.5$ months.
Observe that the calibrated $\rho$-model maintains an overlap higher than $97 \%$ over 9 and 4 months for
Sys.~I and II respectively. The uncalibrated $\rho$-model
performs comparably to the EOB model using the flux of
Ref.~\cite{Gair:2005ih} which has $45$ calibration coefficients, 
both of which have an overlap higher than $97
\%$ over 6 and $1$ month for Sys.~I and II respectively. In the case
of Sys.~II the calibrated flux of Ref.~\cite{Gair:2005ih} perform
better than the uncalibrated $\rho_{\ell m}$ model. Also observe that
the uncalibrated Taylor model is simply inadequate to model EMRIs for
any observation time.

We then see that the use of the calibrated EOB model allows us to
integrate over longer observation times, compared to a 2-week period
or to other models. In turn, this allows us to recover a higher SNR
that we would otherwise. The increase in SNR scales as the square root
of the observation time, as expected. For example, since the calibrated EOB model
is accurate over $9$ months, one would be able to coherently recover
an SNR of $64$ at 1 Gpc for Sys.~I, to be compared with an SNR of
$49$ obtained after $6$ months of coherent integration if using for
example the fluxes of Ref.~\cite{Gair:2005ih}. In general, an
integration over a period longer than two weeks gains us a large
increase in SNR. Such gains in SNR are important because they allow us
to see EMRIs farther out. Since the SNR scales as $\rho \sim
D_{L}^{-1}$, where $D_{L}$ is the luminosity distance, even an SNR
increase in a factor of $3$ increases our accessible volume by a
factor of $27$, since the later scales as $D_{L}^{3}$.

%-----------------------------------------------------------------------------
\section{Higher-Order Effects}
\label{sec:hi-order}

Let us now discuss how finite mass-ratio higher-order effects affect
the GW phase and amplitude.  Those effects are encoded in the $\nu$
terms present in the radiation-reaction force and in the
Hamiltonian. The former are second-order effects in the dissipative
dynamics, while the latter are first-order effects in the conservative
dynamics. We have analytic control over the PN version of such effects
within the EOB formalism, but until now, we had set both of these to
zero when comparing to Teukolsky evolutions, as the latter do not
account for such effects. In the EOB model, however, it is
straightforward to include such terms, as PN expansion are formally
known for all mass ratios at some given order in $v$.  The inclusion
of conservative $\nu$ terms is achieved by using the spin EOB
Hamiltonian of Ref.~\cite{Barausse:2009xi} reviewed in
Appendix~\ref{app:seob}. The inclusion of dissipative $\nu$ terms is
achieved by including relative $\nu$ terms in the multipolarly
decomposed waveform $h_{\ell m}$ and flux ${\cal F}$.

Whether such mass-ratio effects matter depends on the EMRI
considered. In Ref.~\cite{Yunes:2009ef}, it was found that such
effects increase the dephasing between EOB and Teukolsky models by one
order of magnitude after a two year evolution for non-spinning EMRIs,
a result consistent with Ref.~\cite{Pound:2007th}. This effect is
greatly amplified when considering spinning
EMRIs. Table~\ref{table:comparison-mass-ratio} compares the effect
that the inclusion of relative $\nu$ terms in the EOB Hamiltonian and
the radiation-reaction force has on the final dephasing after a
one-year evolution.
\begin{table}
\begin{tabular}{ccccccccc}
\hline\hline
System  & I & I &  I  &  II &  II &  II \\
$q_{1}$ & $-0.9$ & $0.0$ & $0.9$ & $-0.9$ & $0.0$ & $0.9$ \\
\hline
No rel.~$\nu$ & 10.04 & 1.60 & 9.36 & 7.63 & 42.21  & 48.39 \\
$\nu$ in H   & 30.38 & 0.083 & 18.96 & 4.38 & 40.35 & 47.13 \\
%$\nu$ in H   & X & 0.082 & 18.96 & X & X & X  \\
$\nu$ in $\dot{E}$   & 19.99 & 5.31 & 4.49 & 7.24 & 41.50 & 46.55 \\
$\nu$ in H and $\dot{E}$   & 40.32 & 6.83  & 14.08 & 3.98 & 39.64 & 45.29 \\
%$\nu$ in H   & 27.20 & 2.43 & 20.66 & 4.76 & 41.02 & 48.76 \\
%$\nu$ in $\dot{E}$   & 19.99 & 5.31 & 4.49 & 7.24 & 41.50 & 46.55 \\
%$\nu$ in H and $\dot{E}$   & 40.32 & 6.83  & 14.08 & 3.98 & 39.64 & 45.29 \\
\hline\hline
\end{tabular}
\caption{\label{table:comparison-mass-ratio} Absolute value of the total dephasing after a $11.5$ moths of evolution. The first row includes no relative $\nu$ contributions in the Hamiltonian or the radiation-reaction force. The second row includes relative $\nu$ terms in the Hamiltonian, while the third row includes such terms in the radiation reaction force.}
\end{table}
In order to read out the effect of such finite mass-ratio terms in the
phasing, one must compare rows two, three and four to the baseline
given in the first row of Table~\ref{table:comparison-mass-ratio}. For
example, the effect of the $\nu$ terms in the Hamiltonian are such as
to increase the dephasing by $27.20 - 10.04 = 17.16$ radians. Observe
that the conservative and dissipative $\nu$ terms usually push the
dephasing in different directions, partially canceling out when both
of them are present. Even then though, the generic effect of
high-order $\nu$ terms is to increase the rate of dephasing by several
tens of radians are a one year evolution. Notice furthermore that the
magnitude of the effect is here not very large because we are
considering circular equatorial orbits.

Finite mass-ratio effects are clearly suppressed when dealing with circular orbits. This is because, for such orbits outside the ISCO, the effect of the conservative self force is simply to shift the waveform from one orbital frequency to another. Thus, from an observational standpoint, such an effect is unmeasurable. Even though the conservative self-force shifts the ISCO, this effect is still degenerate with a shift of the system's mass parameters. This discussion, however, neglects radiation-reaction, which is crucial to model a true inspiral waveform. One can think of the radiation-reaction force as defining a trajectory through the sequence of orbital energies that an orbit follows.  There is gauge invariant information in this sequence, in the sense that the mapping between energies and orbital frequencies depends on the details of the orbit at each energy level.  For a given radiation-reaction force, the sequence of geodesic orbits (and hence the sequence of frequencies) depends on whether the conservative self-force is included or not. 

As is clear from this discussion, the ``real'' (gauge-invariant) effect of the conservative self-force on quasi-circular inspiral waveforms can be rather subtle. A robust effect, however, does arise if the  self-force acts on a more generic orbit, such as an eccentric one. In that case, this force will act separately on the radial and the azimuthal orbital frequencies, which can leave a potentially strong imprint in the waveform. In principle, even for an inclined circular orbit there could be a strong imprint. In practice, however, the azimuthal and polar orbital frequencies are quite similar, which suggests that perhaps, even in this case, the self-force effects will be small.

With all of this in mind, let us discuss the results presented in Table~\ref{table:comparison-mass-ratio} in more detail.
Our study suggests that the overall effect of  
$\nu$ terms in both $H$ and ${\cal F}$ leads to only $5.2$ and $2.5$
additional radians of phase for non-spinning, Sys.~I and II respectively. This is in
fact consistent with the results presented in
Ref.~\cite{Yunes:2009ef}, except that there one considered $2$-year
long evolutions. One might wonder whether using the non-spinning 
Hamiltonian of Ref.~\cite{Buonanno:2009qa} (where the deformed-Schwarzschild potential 
are Pad\`e-resummed instead of being given by Eqs.~(\ref{eq:D}), (\ref{Deltat})) 
has an effect on this dephasing. We have investigated this question and found
that the additional contribution to the phase is $0.05$ ($0.03$) and
$1.06$ ($1.33$) radians for Sys.~I and II respectively over the entire
year of inspiral using the 3PN (4PN) Pade form of the
deformed potentials. [We notice~\cite{Favata:2010yd} that the 
4PN Pad\`e potentials of Ref.~\cite{Buonanno:2009qa} 
reproduce very closely the ISCO-shift of Ref.~\cite{Barack:2009ey}.] This implies that 
the non-spinning Hamiltonian~\cite{Buonanno:2009qa} at 3PN and 4PN order
is sufficiently close to the Hamiltonian presented in Appendix~B for
data analysis of non-spinning EMRIs.

One can also compare the results in Table~\ref{table:comparison-mass-ratio} to the recent study of Huerta and Gair~\cite{2009PhRvD..79h4021H}, who investigated the effect of $\nu^{2}$-corrections in the determination of parameters, given an EMRI signal. Their Table~I presents the number of cycles accumulated for a variety of mass ratios. Their last column happens to correspond to our Sys.~II with no spin, for which they get a total dephasing of $2.3$ rads and $3.8$ rads after the last year of inspiral when including only conservative and all second-order corrections. This is to be compared to our results: $1.86$ rads in and $2.6$ rads after the last year of inspiral when including only conservative and all second-order corrections. These numbers are in excellent agreement, allowing for differences in the modeling. Their analysis suggests that such small difference will not affect parameter estimation for EMRIs similar to Sys II. For Sys I, however, the dephasing is much larger as the mass ratio is less extreme by one order of magnitude; thus, parameter estimation might be affected in this case.

Another high-order effect that one can study is the inclusion of the
small object's spin in the evolution of the binary. Since the spin
angular momentum of the small body scales with its mass, its
contribution to the orbital evolution is one order in $\nu$
suppressed. We can model this effect by allowing $q_{2} \neq 0$ in the
EOB Hamiltonian of Ref.~\cite{Barausse:2009xi} and letting $\nu \neq
0$. Doing so for Sys.~I (Sys.~II) with $q_{1} = 0.9$ and $q_{2} =
0.75$, we find that the total dephasing now becomes $17.26$ ($45.49$
rads), as shown in Table~\ref{table:comparison-spin}. 
This is to be compared to the case when $q_{2} = 0$, which
returns a dephasing of $14.08$ rads ($45.29$ rads). Thus, the effect
of the second spin contributes roughly $3.3$ rads in this case. 
\begin{table}
\begin{tabular}{ccccccccc}
\hline\hline
System  & I &  I & I & II & II & II \\
$q_{1}$ & $-0.9$ & $0.0$ & $0.9$ & $-0.9$ & $0.0$ & $0.9$ \\
\hline
$q_{2} = -0.75$  & 36.34 & 3.96 & 16.03 & 4.41 & 40.41& 47.04  \\
$q_{2} = 0.00$   & 40.32 & 6.83 & 14.08 & 3.98 & 39.64 & 45.29  \\
$q_{2} = 0.75$   & 44.31 & 9.70 & 12.14 & 3.56 & 38.88 & 43.54  \\
\hline\hline
\end{tabular}
\caption{\label{table:comparison-spin}
Absolute value of the total dephasing in rads after a $11.5$ moths of evolution. The first row sets the second BH's spin to zero, while
the second row sets it to $q_{2} = 0.75$. In both cases, the spin of the background is set to $q_{1} = 0.9$
and $\nu \neq 0$ in both the Hamiltonian and the radiation-reaction force.}
\end{table}
We can also compare these results to those estimated by Barack and Cutler~\cite{Barack:2003fp}. In their Appendix~C, they estimate that for quasi-circular inspirals similar to our Sys.~II, the spin of the second body should induce a dephasing of roughly $1-10$ rads. This is in good agreement with the results corresponding to Sys.~II in our Table~\ref{table:comparison-spin}. Our results are also in good agreement with an upcoming and independent investigation of spin-effects in the PN phasing~\cite{Marc:prep}.

Finally, taking into account the results of including finite mass-ratio 
effects, we can conclude that unless these are precisely modeled, 
it is not worth requiring an agreement better than $10$--$30$ rads when calibrating 
the phase of the test-particle EOB waveforms against the Teukolsky-based waveforms. 

%-----------------------------------------------------------------------------
\section{Conclusions}
\label{sec:conclusions}

We have constructed an EOB model for EMRIs in quasi-circular, equatorial orbits about spinning backgrounds. In the test-particle limit, this model consists of adiabatically evolving a test-particle in the Kerr spacetime using the factorized energy flux of Refs.~\cite{Damour:2008gu, Pan:2010hz}, augmented by 8 calibration coefficients. The latter are determined by comparing the factorized energy flux to a Teukolsky-based flux, built from solutions to the Teukolsky equation in the radiative approximation. In the adiabatic approximation, the EOB waveforms can be constructed in CPU seconds at a very low computational cost. When finite mass-ratio effects and the small object's spin are included, we build the EOB model by numerically solving the Hamilton equations with the spin EOB Hamiltonian of Ref.~\cite{Barack:2003fp} and the Teukolsky-calibrated factorized energy flux augmented by 
finite mass-ratio effects~\cite{Pan:2010hz}.

For both EMRI systems considered, we find excellent phase and amplitude agreement, with dephasing less than one radian, over periods of months. The exact length of the agreement depends on how relativistic the EMRI system is. We also calculated the overlap between EOB and Teukolsky-based waveforms to find it higher than $97 \, \%$ over $4$ to $9$ months, depending on the EMRI system considered. 

The EOB waveforms built here have higher overlaps and better phase agreements that all currently known EMRI models for spinning, equatorial systems, while requiring much fewer calibration parameters. In particular, the EOB model with 8 calibration coefficients outperforms by almost an order of magnitude the numerical kludge waveforms with the calibrated fluxes of Ref.~\cite{Gair:2005ih} which use 45 calibration coefficients. This implies that EOB waveforms with 8 calibration coefficients can be used for longer coherent integrations, allowing us to obtain a $50 \, \%$ increase in SNR. In turn, this implies that EOB waveforms can see EMRIs that are farther out, increasing the accessible volume by at least a factor of two relative to numerical kludge waveforms. Furthermore if we were using 16 calibration coefficients, we could improve the dephasing from $0.91$ rads ($8.7$ rads) to $0.85$ rads ($4.2$ rads) for System I (System II) after 6 months of evolution. In turn, this would decrease the mismatch from $0.2 \%$ ($12 \%$) to $0.19 \%$ ($3.9\%$) for System I and II after $6$ months of evolution. 

Another possible avenue for future research is the calculation of high
PN order terms in the energy flux and waveforms. Our EOB model
relies on the use of accurate fluxes, but for spinning systems, the
flux to infinity is only known up to $4$PN order in the test-particle
limit. This is in contrast to the non-spinning terms that are known to
$5.5$PN order or the BH absorption terms that are known to $6.5$PN
order. The calculation of the spin-dependent terms in the flux to
infinity to $4.5$, $5$ and $5.5$PN order terms in the test-particle
limit is not quixotic and would be invaluable. Once these coefficients
are known, then presumably the EOB waveforms would be more accurate
and might require less adjustable parameters.

Of course, the EOB waveforms we constructed here are less powerful than kludge waveforms~\cite{Barack:2003fp,Babak:2006uv,Gair:2005ih} in their generality. Our waveforms cannot yet model inclined or eccentric inspirals. The inclusion of inclined orbit should be relatively straightforward, but the addition of eccentricity might require some revamping of the EOB framework. Future work in this direction would be definitely worthwhile.

Ultimately, one would like to obtain a waveform model that is sufficiently fast, efficient and accurate to do realistic EMRI data-analysis for LISA. Such a model would need to include the correct self-force contributions to the conservative dynamics, the appropriate second-order terms in the radiation-reaction force and the correct terms that describe the small object's spin. The EOB model we developed here does agree with all known PN self-force calculation to date. However, since full self-force calculations are not yet completed, we do not have a way of assessing the error of including the currently known EOB finite 
mass-ratio effects. In fact, so far  the only comparisons between the PN/EOB and self-force results have been concerned with the non-spinning case, and have been limited to the ISCO shift~\cite{Barack:2009ey,Damour:2009sm} and other gauge invariant quantities~\cite{Blanchet:2009sd,Blanchet:2010zd}. Quite interestingly, the calibration of the EOB model to comparable-mass numerical-relativity simulations improves the agreement of the model to self-force results~\cite{Damour:2009sm,Favata:2010yd}. Thus, we hope that future calibrations of the spin EOB Hamiltonian to comparable-mass simulations of spinning BHs will allow us to build an EOB model which includes finite mass-ratio effects in a more accurate way.

All that said, we have found that the presence of PN self-force terms in the spin EOB model of Ref.~\cite{Barausse:2009xi} leads to dephasing of $10$--$30$ rad over one year depending on the EMRI system and the spin of the background. By contrast, the inclusion of the small object's spin introduces dephasing on the order of a few radians. Taking into account those results, we can conclude that unless those 
finite mass-ratio effects are precisely modeled, currently, it is not worth  requiring an agreement better than $10$--$30$ rads when calibrating the phase of the test-particle EOB waveforms against the Teukolsky-based waveforms. This in turn implies that calibrating $16$ parameters instead of $8$ to an EOB model is overkill as other systematics (induced by neglecting self-force effects) will be dominant.

%-----------------------------------------------------------------------------
\begin{acknowledgments}
We are grateful to Leor Barack, Carlos Sopuerta 
and Frans Pretorius for useful suggestions and comments. 
NY, AB, EB and YP, and SAH acknowledge support from the NSF grants PHY-0745779, 
PHY-0903631, and PHY-0449884; AB, SAH and MCM also acknowledge 
support from NASA grants NNX09AI81G, NNX08AL42G and NNX08AH29G.
NY acknowledges support from the National Aeronautics and Space Administration through Einstein Postdoctoral Fellowship Award Number PF0-110080 issued by the Chandra X-ray Observatory Center, which is operated by the Smithsonian Astrophysical Observatory for and on behalf of the National Aeronautics Space Administration under contract NAS8-03060.
\end{acknowledgments}

%-----------------------------------------------------------------------------
\appendix

\section{Mapping between Spheroidal and Spherical Harmonics}
\label{app:sph}

\subsection{Quantities at Spatial Infinity}

Let us first consider the GW fluxes that describe radiation escaping to infinity, and later discuss the radiation absorbed by the BH. Frequency-domain Teukolsky equation codes expand the curvature scalar $\psi_4$ as
\begin{equation}
\psi_4 = \frac{1}{r}\sum_{\ell m} Z^H_{\ell m} S^{-}_{\ell m}(\theta,\phi)
e^{-i\omega_m t}\;.
\label{eq:psi4expand}
\end{equation}
We have here incorporated the $e^{im\phi}$ dependence into the spin-weighted spheroidal harmonics $S^{-}_{\ell m}$. These harmonics depend on the value of $m_{1} q_{1} \omega_m$, and the minus superscript is a reminder that the we consider here harmonics of spin-weight $-2$.  Throughout this appendix, the index $\ell$ refers to the spheroidal harmonic index, while $l$ refers to the spherical harmonic index.

From $\psi_4$, we compute waveforms via  $\psi_4 = (\ddot h_+ - i \ddot h_\times)/2$, and hence, for a frequency-domain application,
\begin{equation}
h_+ - i h_\times = -\frac{2}{r} \sum_{\ell m}
\frac{Z^H_{\ell m}}{\omega_m^2} S^{-}_{\ell m}(\theta, \phi)
e^{-i\omega_m t}\;.
\label{eq:hexpand}
\end{equation}
We have implicitly assumed that $\omega_m$ is time-independent, or at least that its time-dependence is subleading.  A better expansion is to re-express things in terms of the accumulated phase, i.e.,~the integral of the frequency $\omega_{m} \equiv \dot{\Phi}_{m}$, namely
\begin{eqnarray}
\psi_4 &=& \frac{1}{r}\sum_{\ell m} Z^H_{\ell m} S^{-}_{\ell m}(\theta, \phi)
e^{-i\Phi_m}
\label{eq:psi4}
\\
h &\equiv& h_+ - i h_\times = -\frac{2}{r} \sum_{\ell m}
\frac{Z^H_{\ell m}}{\omega_m^2} S^{-}_{\ell m}(\theta, \phi) e^{-i\Phi_m}\;.
\label{eq:h}
\end{eqnarray}

The EOB and numerical relativity (NR) communities like to project these quantities onto a basis of spin-weighted spherical harmonics. For $\psi_4$, they define $C_{lm}(t,r)$ via
\begin{equation}
\psi_4 = \frac{1}{r}\sum_{lm} C_{lm}(t,r) Y^{-}_{lm}(\theta,\phi)\;.
\label{eq:Clmdefine}
\end{equation}
and the harmonically-decomposed waveforms $h_{lm}(t,r)$ via
\begin{equation}
h = \frac{1}{r}\sum_{lm} h_{lm}(t,r) Y^{-}_{lm}(\theta,\phi)\;.
\label{eq:hlmdefine}
\end{equation}
The minus superscript again denotes spin-weight $-2$. Defining the inner-product
\begin{equation}
\left\langle Y^{-}_{lm} | f \right\rangle = \int d\Omega\,
Y^{-,*}_{lm}(\theta,\phi) f\;,
\end{equation}
the extraction of the $C_{lm}$ and $h_{lm}$ is simple:
\begin{equation}
C_{lm}(t,r) = r \left\langle Y^{-}_{lm} | \psi_4 \right \rangle\;,
\quad
h_{lm}(t,r) = r \left\langle Y^{-}_{lm} | h \right \rangle\;.
\label{eq:Chlminner}
\end{equation}

Let us now take the Schwarzschild limit to see whether these expressions simplify. In that limit
\begin{equation}
S^{-}_{\ell m}(\theta, \phi) = Y^{-}_{\ell m}(\theta,\phi)\;,
\label{eq:spheroidallimit}
\end{equation}
and thus, performing the necessary projections and taking advantage of the orthonormality of spherical harmonics, we find
\begin{eqnarray}
C_{lm}(t,r) &=& Z^H_{lm} e^{-i\Phi_m}\;;
\label{eq:ClmSchw}
\\
h_{lm}(t,r) &=& -\frac{2 Z^H_{lm}}{\omega_m^2}e^{-i\Phi_m}\;.
\end{eqnarray}

In spinning backgrounds, however, the mapping is more complicated. We can use the fact that spheroidal harmonics can be expressed as a sum of spherical harmonics via
\begin{equation}
S^{-}_{\ell m}(\theta, \phi) = \sum_j b^\ell_j \, Y^{-}_{jm}(\theta, \phi)\;.
\label{eq:S-2}
\end{equation}
The dependence on $m_{1} q_{1} \,\omega_m$ enters through the expansion coefficients $b^\ell_j$ (see, e.g. Ref.~\cite{Hughes:2000ssa}). Inserting this expansion into Eq.\ (\ref{eq:psi4expand}) and using the inner-product definition, Eq.~\eqref{eq:Chlminner}, we find
\begin{eqnarray}
C_{jm} &=& e^{-i\Phi_m}\sum_{\ell} b^\ell_j Z^H_{\ell m}\;;
\label{eq:Cjm}
\\
h_{jm} &=& -\frac{2e^{-i\Phi_m}}{\omega_m^2}\sum_{\ell}b^\ell_jZ^H_{\ell m}
= -\frac{2C_{jm}}{\omega_m^2}\;.
\end{eqnarray}
In the Schwarzschild limit, $b^\ell_j = \delta_{\ell j}$, so that Kerr simply limits as it should.

From the definition of the Isaacson stress-energy tensor, one can easily show that
\begin{eqnarray}
\frac{d^2E^\infty}{dtd\Omega} &=& \sum_{\ell m}S^{-}_{\ell m}(\theta,\phi)^2
\frac{|Z^H_{\ell m}|^2}{4\pi\omega_m^2}\;,
\\
&=& \sum_{lm}Y^{-}_{lm}(\theta,\phi)^2
\frac{|C^H_{lm}|^2}{4\pi\omega_m^2}\;.
\end{eqnarray}
We have here used the orthonormality of both the spheroidal and spherical harmonics to simplify the sums, as well as the fact that the time dependence of the $C^H_{lm}$ disappears when its modulus is computed.  Performing the angular integrals leaves us with familiar formulas:
\begin{equation}
\dot E^\infty = \sum_{\ell m} \frac{|Z^H_{\ell m}|^2}{4\pi\omega_m^2}
= \sum_{lm} \frac{|C^H_{lm}|^2}{4\pi\omega_m^2}\;.
\end{equation}
This breaks down nicely enough that it is useful and sensible to define the modal contributions $\dot E_{\ell m}$.

%----------------------------------------------------------------------------------------------------
\subsection{Quantities at Event Horizons}

%This very nicely allows us to
%get flux both to infinity and down the horizon with a single $s = -2$
%code.  Unfortunately, this trick does not generalize to spherical
%harmonics quite as neatly, so we need to include information about the
%$+2$ spin-weight harmonics in this calculation.
%A neat trick allows us to compute the flux down the horizon using a calculation involving spin-weight $-2$ spheroidal harmonics.  

As a general principle, computing quantities that are related to an event horizon is usually more complicated than computing the same quantities at spatial infinity.  For the fluxes, for example, this is because there is no simple generalization of the Isaacson tensor on the horizon.  Instead, one must examine the shear of the horizon's generators, look at how this shear generates entropy, and then apply the area theorem to compute fluxes~\cite{1974ApJ...193..443T}. The relevant quantities at the horizon depend on the Newman-Penrose scalar $\psi_0$, instead of $\psi_{4}$, the former of which is a quantity of spin-weight $+2$, rather than $-2$.  

The GW energy flux per unit solid angle at the horizon is given by
\begin{equation}
\frac{d^2E^H}{dt d\Omega} = \frac{\omega_m m_{1} r_+}{2\pi p_m}|\sigma^{\rm
  HH}|^2\;,
\end{equation}
where $p_m = \omega_m - m\omega_+$, and where $\omega_+ = q_{1}/(2 r_+)$ is the angular velocity of observers co-rotating with the event horizon.  The quantity $\sigma^{\rm HH}$ is the shear to the horizon's generators as found by Ref.~\cite{1972CMaPh..27..283H}. This quantity is fairly simply related to $\psi_0$, so let us introduce an expansion of $\psi_0$ in spin-weight $+2$ spheroidal harmonics
\begin{equation}
\psi_0 = \Delta^{-2}\sum_{\ell m} W^\infty_{\ell m}(r) \, 
S^{+}_{lm}(\theta,\phi) \, e^{-i\Phi_m}\;,
\end{equation}
where $\Delta$ is given in Eq.~\eqref{Delta}. Notice that this quantity diverges on the event horizon $r_{+}$ because the Kinnersley tetrad, which is used to define the projection for $\psi_0$, is ill-behaved as $r \to r_+$.  This can be corrected for by converting to $\sigma^{\rm HH}$ for any given $(\ell,m)$ mode~\cite{1974ApJ...193..443T}
\begin{equation}
\sigma^{\rm HH}_{\ell m} = \Delta^2\gamma_m \psi_{0,\ell m}
= \sum_{\ell m} \gamma_m W^\infty_{\ell m}
S^{+}_{lm}(\theta,\phi)e^{-i\Phi_m}\;.
\end{equation}
The complex number $\gamma_m$ is given by $\gamma_m = -[4(ip_m + 2\epsilon)(2 m_{1} r_+)^2]^{-1}$, where $\epsilon = \sqrt{m_{1}^2 - m_{1}^{2} q_{1}^2}/(4 m_{1} r_+)$.

With this in hand, the GW energy flux at the horizon becomes
\begin{eqnarray}
\frac{d^2E^H}{dt d\Omega} &=& \sum_{\ell m} \frac{\omega_m m_{1} r_+}{2\pi p_m}
|\gamma_m|^2 |W^\infty_{\ell m}|^2 \left(S^{+}_{\ell m}\right)^2
\\ \nonumber 
&=& \sum_{\ell m} \left(S^{+}_{\ell m}\right)^2
\frac{\omega_m^3}{16p_m(p_m^2 + 4\epsilon^2)(2m_{1}r_+)^3}
\frac{|W^\infty_{\ell m}|^2}{4\pi\omega_m^2}\;,
\end{eqnarray}
which integrates to
\begin{equation}
\dot E^H = \sum_{\ell m}\frac{\omega_m^3}{16p_m(p_m^2 +
  4\epsilon^2)(2m_{1}r_+)^3}\frac{|W^\infty_{lm}|^2}{4\pi\omega_m^2}\;.
\end{equation}
Implementing this equation is difficult because it requires that one computes both $\psi_0$ and $\psi_4$ when solving the Teukolsky equation. Since these quantities have different angular dependence and a different source function, this would be a non-trivial undertaking.

Instead, one can take advantage of a remarkable simplification, the so-called Starobinsky identities~\cite{1973ZhETF..65....3S}, that relates $\psi_4$ to $\psi_{0}$ and vice-versa via an algebraic relation. We can use this to relate the coefficients $W^\infty_{\ell m}$ to the coefficients $Z^\infty_{\ell m}$, namely
\begin{equation}
W^\infty_{\ell m} = \beta_{\ell m} Z^\infty_{\ell m}\;,
\label{eq:scidentity}
\end{equation}
where
\begin{eqnarray}
\beta_{\ell m} &=& \frac{64(2m_{1}r_+)^4ip_m(p_m^2 + 4\epsilon^2)(-ip_m +
  4\epsilon)}{c_{\ell m}} Z^\infty_{\ell m}\;,
\nonumber \\
|c_{\ell m}|^2 &=& \left[(\lambda + 2)^2 + 4 q_{1} m_{1} \omega_m - 4 q_{1}^2 m_{1}^{2}\omega_m^2\right] \left(\lambda^2  
\right. 
\nonumber \\
&+& \left.  36 m q_{1} m_{1} \omega_m - 36q_{1}^2 m_{1}^{2} \omega_m^2\right) +\,(2\lambda + 3) 
\nonumber \\
&\times& (96q_{1}^2 m_{1}^{2} \omega_m^2 - 48 m q_{1} m_{1}\omega_m) 
\nonumber \\
&+& 144\omega_m^2 m_{1}^{2} ( 1 - q_{1}^2)\;,
\nonumber \\
{\rm Im}\,c_{\ell m} &=& 12 m_{1} \omega\;,
\nonumber \\
{\rm Re}\,c_{\ell m} &=& +\sqrt{|c_{\ell m}|^2 - 144m_{1}^2\omega^2}\;,
\nonumber \\
\lambda &=& {\cal E}_{\ell m} - 2 q_{1} m_{1} m\omega + q_{1}^2 m_{1}^{2} \omega^2 - s(s+1)\;,
\end{eqnarray}
and ${\cal E}_{\ell m}$ is the eigenvalue of $S^{-}_{\ell  m}(\theta,\phi)$.  

We can then finally write the energy flux formula as
\begin{equation}
\dot E^H = \sum_{\ell m}\alpha_{\ell m}\frac{|Z^\infty_{\ell
    m}|^2}{4\pi\omega_m^2}\;,
\end{equation}
where the coefficient $\alpha_{\ell m}$, given explicitly in Ref.~\cite{Hughes:2000ssa}, agglomerates the factors $\beta$, $\gamma$ into one big expression. This expression reorganizes terms slightly in order for its structure to resemble the expression for the flux to infinity as much as possible. Using this, it is simple to write the modal contribution as decomposed into a spheroidal harmonic basis: $\dot E^H_{\ell m} = \alpha_{\ell m} |Z^\infty_{\ell m}|^2/(4\pi\omega_m^2)$.  

Decomposing the horizon energy flux formula into a spherical harmonic basis is slightly more difficult. The key confusing issue is that there are now {\it two} spherical harmonic basis to worry about: one for each spin-weight.  With respect to these two bases, we can define two spheroidal harmonic expansions:
\begin{eqnarray}
S^-_{\ell m}(\theta,\phi) &=& \sum_j b^\ell_j Y^-_{jm}(\theta,\phi)\;,
\\
S^+_{\ell m}(\theta,\phi) &=& \sum_j d^\ell_j Y^+_{jm}(\theta,\phi)\;.
\end{eqnarray}
(The coefficients $d^\ell_j$ expand the $+2$ spin-weight spheroidal harmonic in $+2$ spin-weight spherical harmonics, just as the coefficients $b^\ell_j$ do so for the $-2$ harmonics.)
It's worth emphasizing that the different spherical harmonics are not simply related to one another.  

The two quantities which can be put into a spherical harmonic basis
are $\psi_0$ and $\psi_4$, both evaluated in the vicinity of the
horizon:
\begin{eqnarray}
\psi_0 &=& \Delta^{-2}\sum_{\ell j m}W^\infty_{\ell m}d^{\ell}_j
Y^+_{jm}(\theta,\phi)e^{-i\Phi_m}
\\
&=& \Delta^{-2}\sum_{jm}U^\infty_{jm} Y^+_{jm}(\theta,\phi)\;,
%\qquad{\rm where} 
\\
U^\infty_{jm} &=& \sum_{\ell} W^\infty_{\ell m}d^{\ell}_je^{-i\Phi_m}\;;
\label{eq:Uinftydefined}
\end{eqnarray}
and
\begin{eqnarray}
\psi_4 &=& \frac{\Delta^2}{(r - i q_{1} m_{1}\cos\theta)^4}\sum_{\ell jm}
Z^\infty_{\ell m}b^{\ell}_j Y^-_{jm}(\theta,\phi)e^{-i\Phi_m}
\\
&=& \frac{\Delta^2}{(r - i q_{1} m_{1}\cos\theta)^4}\sum_{jm}C^\infty_{jm}
Y^-_{jm}(\theta,\phi)\,, 
%\quad{\rm where} 
%\qquad \quad
\\
C^\infty_{jm} &=& \sum_{\ell} Z^\infty_{\ell m}b^{\ell}_je^{-i\Phi_m}\;.
\end{eqnarray}
Using the results presented in this appendix, one can easily find an expression for $\dot E^H$ in terms of the $+2$ harmonic coefficients $U^\infty_{lm}$:
\begin{equation}
\dot E^H = \sum_{lm}\frac{\omega_m^3}{16p_m(p_m^2 +
  4\epsilon^2)(2m_{1}r_+)^3}\frac{|U^\infty_{lm}|^2}{4\pi\omega_m^2}\;.
\label{eq:EdotH_spherical}
\end{equation}
Unfortunately, this is not that useful, as it requires knowledge of the $\psi_0$ expansion coefficients $W^\infty_{\ell m}$.

Taking advantage of the Starobinsky identity again, we can combine Eqs.\ (\ref{eq:scidentity}) and
(\ref{eq:Uinftydefined}) to find 
\begin{eqnarray}
U^\infty_{jm} = \sum_{\ell} \beta_{\ell m}Z^\infty_{\ell m}
d^{\ell}_je^{-i\Phi_m}\;.
\end{eqnarray}
It is then a straightforward to insert this into Eq.\ (\ref{eq:EdotH_spherical}) to obtain the down-horizon flux expanded into modes of $+2$ spherical harmonics.

%-----------------------------------------------------------------------------------------------
\subsection{Truncation issues}

We have so far been rather schematic regarding the limits on all sums. In principle, all these sums should be carried out from some lower limit $l_{\rm min}$ to infinity, where the former is given by $l_{\rm min} = {\rm min}(|s|,|m|)$.  In a numerical application, the upper limit must be truncated at some finite value $l_{\rm max}$. We typically find that the magnitude of terms falls off as a power of $l$. When decomposing into spheroidal harmonics, it is thus typically sufficient  to pick some cutoff value and apply it uniformly. 

Applying such a cutoff is slightly more complicated when we convert to spherical harmonics.  The reason is that a given spheroidal harmonic $\ell$ has contributions from spherical harmonics at index $j > \ell$. Consider, as a concrete example, the spheroidal harmonic $S^{-}_{54}$ for $a/m_{1} = 0.99$, $\omega = 0.1$: the expansion coefficients for this harmonic are
\begin{eqnarray}
b^5_3 &=& -0.0110657\,, \quad b^5_4 = 0.99987\,, \quad b^5_5 = 0.0117368,
\nonumber \\
b^5_6 &=& 0.000123221\,,\quad b^5_7 = 9.4336\times 10^{-7},
\nonumber \\
b^5_8 &=& 6.4276\times 10^{-8}\, \quad b^5_9 = 3.70511\times 10^{-11}\,,
\nonumber \\
b^5_{10} &=& 1.93317\times 10^{-13}\, \quad b^5_{11} = 8.97558\times 10^{-16}\;.
\end{eqnarray}
Coefficients beyond $b^5_{11}$ are small enough that our code does not compute them in this case.  Notice that as we move away from the $j = \ell$ term (whose value is close to unity) the coefficients fall off by roughly powers of $\epsilon \simeq 0.01$.  This behavior is typical, although the value of $\epsilon$ depends strongly on $q_{1}$ and $\omega$ (e.g., $\epsilon \simeq 10^{-3}$ for $q_{1} = 0.1$, $\omega = 0.1$, but $\epsilon \simeq 0.4$ for $q_{1} = 0.999$, $\omega = 5$).

When one converts from $Z^{H}_{\ell m}$ to $C_{jm}$, this behavior forces us to include a {\it buffer} of $\ell$-values beyond the maximum spherical harmonic that we want to compute.  The size of the buffer depends (rather strongly) on the values of $q_{1}$ and $\omega$. In the weak field, where $q_{1} m_{1} \omega \ll 1$ even for large $m$, it is enough to include a buffer of $2$ (i.e., such that $\ell_{\rm max} = j_{\rm max} + 2$). For the results used in this paper, we have chosen a uniform buffer region of size $8$.
  
%-----------------------------------------------------------------------------
\section{GW Energy Absorption}
\label{app:flux-abs}

The $N/2$-th order Taylor expansion of the flux in PN theory is given by the series~\cite{lrr-2005-3}
\be
{\cal{F}}_{\Ta}^{(N)} = {\cal{F}}_{\rm Newt.} \sum_{n=0}^{N} \left[a_{n}(\nu) + b_{n}(\nu) \log(v)\right] v^{n},
\label{Taylor-flux}
\ee
where ${\cal{F}}_{\rm Newt} \equiv 32/5 \nu^{2} v^{10}$ is the leading-order (Newtonian) piece of the flux, $v$ is the circular orbital frequency, $\log$ stands for the natural logarithm while $a_{n}$ and $b_{n}$ are PN parameters, with $b_{n<6} = 0$. 

The PN parameters can be classified according to their physical origin and whether they include spin contributions or not. The flux pieces that account for GW emission to infinity are well-known and, for example, are given in Ref.~\cite{Boyle:2008ge,Arun:2008kb}. Those that correspond to radiation in-falling into the horizon will be labeled $(a^{\Ho}_{n},b^{\Ho}_{n})$, with a superscript $S$ ($NS$) if they are further spin-dependent (spin-independent).

The coefficients associated with BH absorption can only be formally obtained employing BH perturbation theory, as PN theory treats BHs as effective test particles.  The logarithm-independent terms associated with non-spinning contributions to the radiation flux through the horizon are
\ba
a_{8}^{\Ho, \NS} &=& 1, \quad
a_{9}^{\Ho, \NS} = 0, \\
a_{10}^{\Ho, \NS} &=& 4, \quad
a_{11}^{\Ho, \NS} = 0, \\
a_{12}^{\Ho, \NS} &=& \frac{172}{7}, \quad
a_{13}^{\Ho, \NS} = 0, \\
\ea
where $a_{<8}^{\Ho, \NS} = 0$. All logarithm-dependent terms identically vanish here: $b_{n}^{\Ho, \NS} = 0$. Similarly, the logarithm-independent terms associated with spinning contributions to the radiation flux through the horizon are
\ba
a_{5}^{\Ho, \Sp} &=& - \frac{\bar{q}}{4} - \frac{3 \bar{q}^{3}}{4}, \\
a_{6}^{\Ho, \Sp} &=& 0, \\
a_{7}^{\Ho, \Sp} &=& - \bar{q} - \frac{33 \bar{q}^{3}}{16}, \\
a_{8}^{\Ho, \Sp} &=& - \frac{1}{2} + \frac{35}{6} \bar{q}^2 - \frac{3}{12} \bar{q}^4 
\nonumber \\
&+& \left(\frac{1}{2} + \frac{13}{2} \bar{q}^2 + 3 \bar{q}^4\right) \left(1 - \bar{q}^2\right)^{1/2} 
\nonumber \\
&+& i \bar{q} \left(1+ 3 \bar{q}^2\right)) \left\{ \psi^{(0)}\left[3 - 2 i \bar{q} \left(1 - \bar{q}^2\right)^{-1/2}\right] 
\right. 
\nonumber \\
&-& \left. \psi^{(0)}\left[3 + 2 i \bar{q} \left(1 - \bar{q}^2\right)^{-1/2}\right] \right\}, \\
a_{9}^{\Ho, \Sp} &=& -\frac{43 \bar{q}}{7} - \frac{4651 \bar{q}^{3}}{336} - \frac{17 \bar{q}^{5}}{56}, \\
a_{10}^{\Ho, \Sp} &=& -2 + \frac{433}{24} \bar{q}^2 - \frac{95}{24} \bar{q}^4 
\nonumber \\
&+& \left(2 + \frac{163}{8} \bar{q}^2 + \frac{33}{4} \bar{q}^4 \right) \left(1 - \bar{q}^{2}\right)^{1/2}
\nonumber \\
&-& \frac{3}{24} i \bar{q} \left(4 - 3 \bar{q}^2\right) \left\{ \psi^{(0)}\left[3 + i \bar{q} \left(1 - \bar{q}^2\right)^{-1/2}\right] 
\right.
\nonumber \\
&-& \left. \psi^{(0)}\left[3 - i \bar{q} \left(1 - \bar{q}^2\right)^{-1/2}\right] \right\}  - 3 i \bar{q} \left(1 + 3 \bar{q}^2\right) 
\nonumber \\
&\times& \psi^{(0)}\left[3 + 2 i \bar{q} \left(1 - \bar{q}^2\right)^{-1/2}\right] + 3 i \bar{q} \left(1 + 3 \bar{q}^2 \right) 
\nonumber \\
&\times& \psi^{(0)}\left[3 - 2 i \bar{q} \left(1 - \bar{q}^2\right)^{-1/2}\right], 
\label{polygamma}
\ea
where $a_{<5}^{\Ho, \Sp} =0$ and where the polygamma function $\psi^{(n)}(z) \equiv (d^{n}\Gamma(z)/dz) \; \Gamma(z)^{-1}$ is the $n$th-derivative of the Gamma function. The coefficients $(a_{11}^{\Ho, \Sp},a_{12}^{\Ho, \Sp})$ are also known, but we do not write them out here as they are lengthy and unilluminating [e.g., see Appendix J in Ref.~\cite{Mino:1997bx}]. Notice that the BH absorption coefficients in the spinning case are non-zero starting at $2.5$ PN order, which is to be contrasted with the non-spinning BH absorption terms that start at $4$ PN order.

An ambiguity exists when incorporating these BH absorption contributions into the flux. As one can observe, the spin-dependent
coefficients $a_{n}^{\rm Hor, S}$ depend on the spin parameter of the background $\bar{q}$, for which one could choose the real spin parameter $\bar{q} = q_{1}$ or the effective spin parameter $\bar{q} = q$, defined in Appendix~\ref{app:seob}. Since $q = q_{1} + {\cal{O}}(m_{2}/m_{1})$, these choices are identical in the test particle limit, when we calibrate to Teukolsky-fluxes. In lack of better guidance, we here choose $\bar{q} = q$.   

%-----------------------------------------------------------------------------
\section{Spin EOB Hamiltonian}
\label{app:seob}

In Sec.~\ref{sec:hi-order} we have investigated how the analytical results calibrated to the Teukolsky-based 
waveforms change when we switch on the PN conservative self-force and the second-order radiation 
reaction effects, and when we include the spin of the small object. This study employed the spin EOB 
Hamiltonian of Ref.~\cite{Barausse:2009xi}, which was derived from Ref.~\cite{Barausse:2009aa}, building also on results of Ref.~\cite{Damour:2008qf}. As we shall review below, the Hamiltonian of Ref.~\cite{Barausse:2009xi} reproduces the known spin-orbit (spin-spin) PN couplings 
through 2.5PN (2PN) order for comparable masses, and {\it all} PN couplings linear in the spin 
of the small object in the test-particle limit. We shall here restrict attention to circular, equatorial orbits, 
and assume that the spins are aligned with the orbital angular momentum.

The motion of a spinning test-particle in a generic curved spacetime 
is described by the Papapetrou equation~\cite{Papa51,Papa51spin,CPapa51spin}. 
Reference~\cite{Barausse:2009aa} derived a Hamiltonian whose Hamilton equations 
are equivalent to the Papapetrou equation. This Hamiltonian therefore describes the motion of a 
spinning particle in a generic curved spacetime, and Ref.~\cite{Barausse:2009xi} computed it in 
the particular case of the Kerr spacetime in Boyer-Lindquist coordinates. 
Denoting with $\mathbf{S}_{1}$ and $m_1$ the spin and mass of the background BH, 
and with $\mathbf{S}_{2}$ and $m_2$ the spin and the mass of the smaller BH, the Hamiltonian 
of a spinning test-particle in Kerr has the generic form~\cite{Barausse:2009xi}
\begin{equation}
H = H_{\rm NS} + H_{\rm S}\,, 
\end{equation}
where $H_{\rm NS}$ is the Hamiltonian of a non-spinning test particle
in Kerr, given by Eq.~(\ref{eq:Hnsdef}), while $H_{\rm S}$ depends on
$\mathbf{S}_{1}$ and $\mathbf{S}_{2}$ and, if PN expanded, generates
{\it all} PN terms linear in the small object's spin $\mathbf{S}_{2}$.

In Ref.~\cite{Barausse:2009xi} the authors constructed the spin EOB Hamiltonian by mapping 
the PN Hamiltonian of two BHs of masses $m_{1,2}$ and spins $\mathbf{S}_{1,2}$ into the 
{\emph{effective}} Hamiltonian of a spinning test-particle of mass $\mu=m_1m_2/(m_1+m_2)$ and spin $\mathbf{S}^\ast$ moving 
in a deformed-Kerr spacetime with mass $M=m_1+m_2$ and spin $\mathbf{S}_{\rm Kerr}$, $\nu = \mu/M$ 
being the deformation parameter. Note that the deformed-Kerr spin parameter 
$q \equiv |\mathbf{S}_{\rm Kerr}|/M^2 \neq q_{1}$, but instead $q \approx q_{1} (1 - 2 m_{2}/m_{1} + \ldots )$ when $m_{2}/m_{1} \ll 1$.

The effective Hamiltonian is~\cite{Barausse:2009xi} 
\begin{eqnarray}
\label{HeffEOBco}
H_{\rm eff} &=& H_{\rm NS}+H_{\rm S}-\frac{\mu}{2M\, r^3}\,\, S_\ast^2\,,
\end{eqnarray}
where $H_{\rm NS}$ is the Hamiltonian of a non-spinning effective particle in the deformed-Kerr background, 
\begin{equation}
H_{\rm NS}=p_{\phi}\,\frac{\widetilde{\omega}_{\rm fd}}{\Lambda_t}+ \frac{\mu\,r\,\sqrt{\Delta_t}\,\sqrt{Q}}{\sqrt{\Lambda_t}}\,,
\label{HNSd}
\end{equation}
which differs from Eq.~(\ref{HKerr}) in that the Kerr potentials $\Delta$ and ${\omega}_{\rm fd}$ have been replaced by their
deformed forms $\Delta_t$ and ${\widetilde{\omega}_{\rm fd}}$ (and also $m_1 \rightarrow M$, $q_1 \rightarrow q$, $m_2\rightarrow \mu$). 
Furthermore, $H_{\rm S}$ in Eq.~(\ref{HeffEOBco}) is linearly proportional to the effective particle's spin $S_{\ast}$ and reads
\begin{widetext}
\begin{eqnarray}
H_{\rm S}&=&\frac{S_\ast}{2 \mu\, M\, \sqrt{\Delta_t}\, \Lambda_t^{5/2}\,\left(\sqrt{Q}+1\right)\,\sqrt{Q}\,r^2}\,
\Bigg\{2\mu\,\sqrt{\Delta_t}\,\Lambda_t\,\left(\sqrt{Q}+1\right) 
\left(\sqrt{\Delta_t}\, p_\phi\, r^3+\mu\,\sqrt{\Lambda_t\, Q}\,\tilde{\omega}_{\rm fd}\right)\,
   r^2\nonumber \\
&& +\sqrt{\Delta_r}\,\Big[\mu\,\Delta_{t_,r}\,\sqrt{\Lambda_t}\, p_\phi\,\left(\left(2 \sqrt{Q}+1\right)\,
(r^2+M^2\,q^2)^2-\Lambda_t\,\left(\sqrt{Q}+1\right)\right)\, r^3 \nonumber \\
&& +2\mu\,\Delta_t\, \sqrt{\Lambda_t}\,
p_\phi\, \left(2 \sqrt{Q}+1\right)\,\left(\Lambda_t-2 r^2\, (r^2+M^2\,q^2)\right)\,r^2 \nonumber \\
&& +\sqrt{\Delta_t}\,\left(p_\phi^2\,r^2+\mu^2 \Lambda_t\,\sqrt{Q}\, \left(1+\sqrt{Q}\right)\right)\,
(\Lambda_t\, \tilde{\omega}_{\rm fd,r}-\Lambda_{t,r}\,\tilde{\omega}_{\rm fd})\Big]\Bigg\}\,,
\label{HS}
\end{eqnarray}
\end{widetext}
where we denote with a comma the derivative with respect to $r$. The 
term proportional to $S_\ast^2$ in Eq.~(\ref{HeffEOBco}) is added to reproduce known spin-spin results at 2PN order. 
The quantities (\ref{HNSd}) and (\ref{HS}) depend on the Kerr-deformed potentials $\Delta_t$, $\Delta_r$, $\Lambda_t$, 
$\widetilde{\omega}_{\rm fd}$, while 
\begin{equation}
Q=1+\frac{p_\phi^2\, r^2}{\mu^2\, \Lambda_t}\,.
\label{Qeob}
\end{equation}
In particular, we have~\cite{Barausse:2009xi} 
\begin{equation}
\label{Lambdat}
\Lambda_t = (r^2 + M^2\,q^2)^2 - M^2\,q^2\,\Delta_t\,, 
\end{equation}
and 
\begin{equation}
\widetilde{\omega}_{\rm fd} = 2 q\, M^2\, r + \omega_1^{\rm fd}\,\nu\,
\frac{q\,M^4}{r} + \omega_2^{\rm fd}\,\nu\,
\frac{q^3\,M^4}{r}\,,
\label{eq:omegaTilde}
\end{equation}
where $\omega_1^{\rm fd}$ and $\omega_2^{\rm fd}$ are two adjustable parameters regulating 
the frame dragging {\it strength}.
Although precise values for $\omega_1^{\rm fd}$ and $\omega_2^{\rm fd}$ can only be determined 
by calibrating the model against NR simulations of comparable-mass spinning BHs, a preliminary 
comparison of the final spin predicted by the EOB model to NR results~\cite{Rezzolla:2007rz,Barausse:2009uz} 
suggests that $\omega_1^{\rm fd}\approx -10$ and $\omega_2^{\rm fd}\approx 20$. 

The deformed-Kerr potential $\Delta_t$ is given at 3PN order by 
\begin{eqnarray}
\label{Deltat}
\Delta_t &=& r^2 \,\left [A(u) + q^2\,u^2 \right ]\,,\\
A(u) &=& 1 - 2\,u + 2 \nu\,u^3 + \nu\,\left (\frac{94}{3} - \frac{41}{32}\,\pi^2 \right )\,u^4\,.
\nonumber \\
\end{eqnarray}
When setting $\nu = 0$, $\Delta_t$ reduces to the Kerr expression 
(\ref{Delta}) (with $m_1 \rightarrow M$, $q_1 \rightarrow q$) $\Delta=\Delta_t= r^2 - 2M\,r + q^2\,M^2$. 
In order to guarantee the presence of deformed horizons (which correspond to the zeros of $\Delta_t$), Ref.~\cite{Barausse:2009xi} suggested 
to rewrite Eq.~(\ref{Deltat}) as ($u \equiv M/r$)
\begin{eqnarray}
\label{Deltatlog}
\Delta_t &=& r^2\, \left [\frac{1-2u\,(1-K\,\nu)}{(1-K\,\nu)^2} + q^2\,u^2\right ]\times\\
&& \left [1 + \nu\,\Delta_0 + \nu \,\log \left (1 + \sum_{i=1}^4 \Delta_i\,u^i \right ) \right ]\,,
\end{eqnarray}
with
\begin{equation}
K = K_0 + 4 (K_1-K_0)\,\nu\,,
\end{equation}
and
\begin{eqnarray}
\Delta_0 &=& K\,(\nu\,K - 2)\label{eq:k_0}\,,\\
\Delta_1 &=& -2(\nu\,K - 1)\,(K+\Delta_0)\,,\\
\Delta_2 &=& \frac{1}{2}\,\Delta_1\, (-4 \nu\,K +\Delta_1 +4)-q^2\,(\nu\,K-1)^2\, \Delta_0\,,\nonumber \\\\
\Delta_3 &=& \frac{1}{3}\,\Big [-\Delta_1^3+ 3 (\nu \,K-1)\, \Delta_1^2+3 \Delta_2\, 
\Delta_1 \nonumber \\
&& -6 (\nu \,K-1) \,
(-\nu \,K + \Delta_2+1) \nonumber \\
&& -3 q^2\, (\nu\,K-1)^2\, \Delta_1\Big ]\,,\\
\Delta_4 &=& \frac{1}{12}\Big\{6 q^2\, \left(\Delta_1^2-2 \Delta_2\right) (\nu\,K-1)^2+ 3 \Delta_1^4
\nonumber \\
&& - 8(\nu\,K-1)\,\Delta_1^3-12 \Delta_2\,\Delta_1^2 \nonumber \\
&& +12 \left [2 (\nu\,K-1)\,\Delta_2+\Delta_3\right]\, \Delta_1  \nonumber\\
&&  +12 \left (\frac{94}{3} - \frac{41}{32} \pi^2\right)\,(\nu\,K-1)^2 \nonumber \\
&& +6 \left [\Delta_2^2-4 \Delta_3 \,(\nu\,K -1)\right ]\Big\}\,,
\label{eq:k_4}
\end{eqnarray}
When expanding Eq.~(\ref{Deltatlog}) through 3PN order, one recovers Eq.~(\ref{Deltat}). 
The quantity (\ref{Deltatlog}) depends on two parameters $K_0$ and $K_1$.  
$K_0$ is fixed to the value $1.4467$ in order to reproduce
the results of Ref.~\cite{Barack:2009ey} for the shift of the ISCO frequency due to the conservative part of the self force.
Also, recent comparisons of the EOB model with numerical simulations of non-spinning comparable mass BHs have suggested 
$K_1 \approx 3/4$. 

The deformed-Kerr potential $\Delta_r$ is given by~\cite{Barausse:2009xi} 
\begin{eqnarray}
\label{eq:D}
\Delta_r &=& \Delta_t\,\left \{1+\log[1 + 6 \nu\, u^2 + 2 (26 - 3 \nu)\, \nu\, u^3] \right \}\,,\nonumber \\
\end{eqnarray}
which reduces to the Kerr-potential $\Delta$ in the limit $\nu =0$ (with $m_1 
\rightarrow M$, $q_1 \rightarrow q$). 

Finally, the spins $\mathbf{S}_{\rm Kerr}$ and $\mathbf{S}^\ast$ in the 
effective description are not equal to $S_{1}$ a $S_{2}$, are instead given by
\begin{eqnarray}
\label{mapping1}
{{S}}^\ast &=& {S}_1\,\frac{m_2}{m_1}+{S}_2\,\frac{m_1}{m_2} +\frac{1}{c^2}\,{\Delta}_{{S}^\ast}\,,\\
\label{mapping2}
{{S}}_{\rm Kerr} &=& {S}_1+{S}_2\,,
\end{eqnarray}
where 
\begin{eqnarray}
{\Delta}_{S^\ast}&=&\frac{\nu}{12}\,  
\left \{ \frac{2M}{r}\,\left [7 \left ({S}_1\,\frac{m_2}{m_1}+{S}_2\,\frac{m_1}{m_2}\right ) 
\right . \right . \nonumber \\
&&  -4 ({S}_1+{S}_2) \bigg] + (Q-1)\, \bigg[3 ({S}_1+{S}_2) 
\nonumber \\
&&\left. \left . +4 \left ({S}_1\,\frac{m_2}{m_1}+{S}_2\,\frac{m_1}{m_2}\right ) \right ]
\right \}\,,
\end{eqnarray}
With all of this at hand, the EOB Hamiltonian used in Sec.~\ref{sec:hi-order} is 
\begin{equation}
\label{hreal}
H_\mathrm{EOB} = M\,\sqrt{1+2\nu\,\left(\frac{H_{\rm eff}}{\mu}-1\right)}\,. 
\end{equation}
A few final observations are due at this point. 
When the smaller BH has zero spin $S_2=0$ and
mass $m_2\ll m_1$, at lowest order in $m_2/ m_1$ the EOB Hamiltonian of Eq.~\eqref{hreal} 
reduces to the Hamiltonian of a non-spinning test particle 
in Kerr. This is because both $S_{\ast}$ and the deformations of the
Kerr potentials are ${\cal O}(m_2/m_1)$. However, at the next-to-leading order in the 
mass-ratio, the EOB Hamiltonian presents corrections with respect to the Hamiltonian of a non-spinning test particle
in Kerr, \textit{(i)} because of the deformations of the Kerr potentials; \textit{(ii)} because of the effective spin $S_{\ast}$, 
which is not zero; \textit{(iii)} because of the higher-order terms in $\nu$ that one obtains expanding Eq.~\eqref{hreal}.  
These corrections encode the conservative part of the self-force in the EOB framework~\cite{Barausse:2009xi}.

\bibliographystyle{apsrev}
\bibliography{review}
\end{document}